\documentclass[11pt]{article}
\usepackage{jheppub} 
 \usepackage[T1]{fontenc}
 \usepackage{booktabs}
 \usepackage{xcolor} 
 \usepackage{xspace}
 \usepackage{dcolumn}
 \usepackage{hyperref}
 \usepackage{caption}
 \usepackage
[subrefformat=parens,position=top,skip=-15pt,margin=15pt,justification=justified,singlelinecheck=false]
 {subcaption}
 \usepackage{placeins}   
 \usepackage[group-digits=false]{siunitx}
 \usepackage{microtype}
 \usepackage{lmodern}
 \usepackage{hyperref}

\pdfoutput=1

\allowdisplaybreaks[1]

\newcommand{\Pl}{\ell}

\newcommand{\fb}{{\ensuremath\unskip\,\text{fb}}\xspace}

\def\refeq#1{\mbox{(\ref{#1})}}
\def\reffi#1{\mbox{Fig.~\ref{#1}}}
\def\reffis#1{\mbox{Figs.~\ref{#1}}}
\def\refta#1{\mbox{Table~\ref{#1}}}

\def\refse#1{\mbox{Section~\ref{#1}}}

\def\citere#1{\mbox{Ref.~\cite{#1}}}
\def\citeres#1{\mbox{Refs.~\cite{#1}}}

\newcommand{\rd}{\mathrm d}

\newcommand{\ie}{\emph{i.e.}\ }
\newcommand{\eg}{\emph{e.g.}\ }

\def\be{\begin{equation}}
\def\ee{\end{equation}}

\newcommand{\PH}{\ensuremath{\text{H}}\xspace}
\newcommand{\Pj}{\ensuremath{\text{j}}\xspace}
\newcommand{\Pp}{\ensuremath{\text{p}}\xspace}
\newcommand{\Pe}{\ensuremath{\text{e}}\xspace}

\newcommand{\Pb}{\ensuremath{\text{b}}\xspace}
\newcommand{\Pq}{\ensuremath{q}\xspace}
\newcommand{\Pt}{\ensuremath{\text{t}}\xspace}
\newcommand{\Pu}{\ensuremath{\text{u}}\xspace}
\newcommand{\Pd}{\ensuremath{\text{d}}\xspace}
\newcommand{\Ps}{\ensuremath{\text{s}}\xspace}
\newcommand{\Pc}{\ensuremath{\text{c}}\xspace}
\newcommand{\Pg}{\ensuremath{\text{g}}\xspace}

\newcommand{\PW}{\ensuremath{\text{W}}\xspace}
\newcommand{\PZ}{\ensuremath{\text{Z}}}

\newcommand{\Mt}{\ensuremath{m_\Pt}\xspace}

\newcommand{\MWOS}{\ensuremath{M_\PW^\text{OS}}\xspace}
\newcommand{\MW}{\ensuremath{M_\PW}\xspace}
\newcommand{\MZOS}{\ensuremath{M_\PZ^\text{OS}}\xspace}
\newcommand{\MZ}{\ensuremath{M_\PZ}\xspace}
\newcommand{\Mb}{\ensuremath{m_\Pb}\xspace}
\newcommand{\Gt}{\ensuremath{\Gamma_\Pt}\xspace}
\newcommand{\GH}{\ensuremath{\Gamma_\PH}\xspace}

\newcommand{\GZOS}{\ensuremath{\Gamma_\PZ^\text{OS}}\xspace}

\newcommand{\GWOS}{\ensuremath{\Gamma_\PW^\text{OS}}\xspace}

\newcommand{\GeV}{\ensuremath{\,\text{GeV}}\xspace}
\newcommand{\TeV}{\ensuremath{\,\text{TeV}}\xspace}

\newcommand{\cw}{c_{\mathrm{w}}}
\newcommand{\sw}{s_{\mathrm{w}}}
\newcommand{\alphas}{\ensuremath{\alpha_\text{s}}\xspace}
\newcommand{\order}[1]{\ensuremath{\mathcal{O}{\left(#1\right)}}\xspace}

\newcommand{\GF}{\ensuremath{G_\mu}}

\newcommand{\ptsub}[1]{\ensuremath{p_{\text{T},#1}}\xspace}

\newcommand{\MVOS}{\ensuremath{M_{V}^\text{OS}}\xspace}%
\newcommand{\GVOS}{\ensuremath{\Gamma_{V}^\text{OS}}\xspace}%

\newcommand{\newc}{\newcommand}
\newc{\bi}{\begin{itemize}}
\newc{\ei}{\end{itemize}}
\newc{\benu}{\begin{enumerate}}
\newc{\eenu}{\end{enumerate}}
\newc{\bc}{\begin{center}}
\newc{\ec}{\end{center}}
\newc{\bfig}{\begin{figure}}
\newc{\efig}{\end{figure}}
\newc{\qbar}{\bar{q}}
\newc{\go}{\tilde{g}}
\newc{\PB}{\textsc{Powheg-Box}}

\newcommand{\recola}{{\sc Recola}\xspace}

\newcommand{\openloops}{O\protect\scalebox{0.8}{PEN}L\protect\scalebox{0.8}{OOPS}\xspace}

\newcommand{\mocanlo}{{\sc MoCaNLO}\xspace}
\newcommand{\mocanlorecola}{{\sc MoCaNLO+Recola}\xspace}
\newcommand{\collier}{{\sc Collier}\xspace}

\newcommand{\madgraph}{{\sc\small MadGraph5\_aMC@NLO}\xspace}

\newcommand{\rT}{{\mathrm{T}}}
\newcolumntype{.}{D{.}{.}{-1}}
\newcolumntype{d}[1]{D{.}{.}{#1}}

\colorlet{tableoverheadcolor}{gray!37.5}
\colorlet{tableheadcolor}{gray!25}
\colorlet{tablerowcolor}{gray!12.5}

\newlength{\width}
\newlength{\height}

\def\draftdate{\relax}
\def\mda{\relax}
\def\mua{\relax}
\def\mla{\relax}
\def\draft{
\def\thtystars{******************************}
\def\sixtystars{\thtystars\thtystars}
\typeout{}
\typeout{\sixtystars**}
\typeout{* Draft mode!
         For final version remove \protect\draft\space in source file *}
\typeout{\sixtystars**}
\typeout{}
\def\draftdate{\today}
\def\mua{\marginpar[\boldmath\hfil$\uparrow$]%
                   {\boldmath$\uparrow$\hfil}\color{black}%
                    \typeout{marginpar: $\uparrow$}\ignorespaces}
\def\mda{\color{red}\marginpar[\boldmath\hfil$\downarrow$]%
                   {\boldmath$\downarrow$\hfil}%
                    \typeout{marginpar: $\downarrow$}\ignorespaces}
\def\mla{\marginpar[\boldmath\hfil$\rightarrow$]%
                   {\boldmath$\leftarrow $\hfil}%
                    \typeout{marginpar: $\leftrightarrow$}\ignorespaces}
\def\Mua{\marginpar[\boldmath\hfil$\Uparrow$]%
                   {\boldmath$\Uparrow$\hfil}\color{black}%
                    \typeout{marginpar: $\uparrow$}\ignorespaces}
\def\Mda{\color{red}\marginpar[\boldmath\hfil$\Downarrow$]%
                   {\boldmath$\Downarrow$\hfil}%
                    \typeout{marginpar: $\downarrow$}\ignorespaces}
\def\Mla{\marginpar[\boldmath\hfil\textcolor{red}{$\Rightarrow$}]%
                   {\boldmath\textcolor{red}{$\Leftarrow $}\hfil}%
                    \typeout{marginpar: $\leftrightarrow$}\ignorespaces}
\overfullrule 5pt
\oddsidemargin 15mm
\marginparwidth 29mm
}

\title{\hfill ~\\[-58mm]
\phantom{h} \hfill\mbox{\small {FR-PHENO-2021-08}, {VBSCAN-PUB-06-21}}
\\[1cm]
\vspace{13mm}   
Full NLO predictions for vector-boson scattering into Z~bosons and its
irreducible background at the LHC} 

\subheader{\today}

\author{Ansgar Denner$^1$,}
\author{Robert Franken$^1$,}
\author{Mathieu Pellen$^2$,}
\author{Timo Schmidt$^1$}

\affiliation{$^1$Universit\"at W\"urzburg, %
        Institut f\"ur Theoretische Physik und Astrophysik, \\ %
        Emil-Hilb-Weg 22,  %
        97074 W\"urzburg, %
        Germany%
}
\affiliation{$^2$Universit\"at Freiburg, %
        Physikalisches Institut, \\ %
        Hermann-Herder-Stra\ss e 3, %
        79104 Freiburg, %
        Germany}
\emailAdd{ansgar.denner@physik.uni-wuerzburg.de}
\emailAdd{robert.franken@physik.uni-wuerzburg.de}
\emailAdd{mathieu.pellen@physik.uni-freiburg.de}
\emailAdd{timo.schmidt@physik.uni-wuerzburg.de}

\abstract{
Vector-boson scattering into two Z~bosons at the LHC is a key channel
for the exploration of the electroweak sector of the Standard Model. 
It allows for the full reconstruction of the scattering process but at the price of a huge irreducible background.
For the first time, we present full next-to-leading-order predictions
for $\Pp\Pp \to \Pe^+\Pe^-\mu^+\mu^-\Pj\Pj+X$ including all
electroweak and QCD contributions for vector-boson scattering signal
and irreducible background.
The results are presented in the form of cross sections and differential distributions.
A particular emphasis is put on the newly computed $\order{\alphas^2 \alpha^5}$ corrections.}

\begin{document}

\maketitle

\newpage

\section{Introduction}

An important part of the physics programme conducted at the Large
Hadron Collider (LHC) aims at extracting Standard Model parameters by measuring precisely particular processes.
This requires the definition of tailored event selections and the subtraction of undesirable background processes.
In the presence of irreducible backgrounds, this picture gets complicated through non-vanishing interferences between the signal and background processes.
For vector-boson scattering (VBS), the irreducible background can be overwhelming which
warrants a detailed analysis of both the signal and background in order to fully exploit its physics potential \cite{Covarelli:2021gyz,1866947}.

The present article continues a series of studies
\cite{Biedermann:2016yds,Biedermann:2017bss,Denner:2019tmn,Denner:2020zit}
aiming at describing not only VBS processes but also their irreducible backgrounds with next-to-leading-order
(NLO) accuracy.  Given that the
signal and background are connected through interferences, a physical
definition of a signal and background separately is not possible (even
when using exclusive cuts to enhance the electroweak (EW) component).
At NLO, the situation becomes even more complicated as a separation of
signal and background in an IR-finite way must necessarily rely on
approximations \cite{Ballestrero:2018anz}.

On top of these considerations, there are more pragmatic reasons to
investigate in detail both the EW and QCD components of VBS at full NLO accuracy.
While for the same-sign W channel the signal-to-background ratio is
about $8/1$, it is only of the order of $1/2$ for the ZZ case.
In addition, given the smallness of the cross sections it is critical to know all contributions, including the loop-induced contribution.

Important steps towards describing $\Pp\Pp \to \Pe^+\Pe^-\mu^+\mu^-\Pj\Pj+X$ at full NLO accuracy have already been taken, especially regarding NLO QCD corrections.
These are known for the EW component \cite{Jager:2006cp} in the VBS approximation and have been implemented in the {\tt POWHEG BOX} framework \cite{Alioli:2010xd} to also include parton-shower (PS) corrections \cite{Jager:2013iza}.
In addition, NLO QCD corrections to the QCD component are known \cite{Campanario:2014ioa} and can in principle be obtained at NLO QCD+PS accuracy from public multi-purpose Monte Carlo generators such as \madgraph \cite{Alwall:2014hca} or {\sc Sherpa} \cite{Sherpa:2019gpd}.
The full NLO corrections of orders $\order{\alpha^7}$ and $\order{\alphas \alpha^6}$ have been computed in \citere{Denner:2020zit} 
along with the loop-induced contribution of order $\order{\alphas^4 \alpha^4}$.
Finally, in \citere{Li:2020nmi}, results for loop-induced ZZ~production with up to 2~jets merged and matched to parton showers have been presented.
Therefore, at NLO level the only missing piece is the order
$\order{\alphas^2 \alpha^5}$, which is for
simplicity sometimes referred to as EW corrections to the QCD component in
the following.
With the present publication we fill this gap by presenting for the first time the full NLO corrections to $\Pp\Pp \to \Pe^+\Pe^-\mu^+\mu^-\Pj\Pj+X$ at the LHC in a unique setup.

Previous works \cite{Biedermann:2016yds,Biedermann:2017bss,Denner:2019tmn,Denner:2020zit} have put a strong emphasis on EW corrections of order $\order{\alpha^7}$.
These are exceptionally large for EW corrections and are an intrinsic feature of VBS at the LHC \cite{Biedermann:2016yds}.
They are even the largest NLO contribution for the same-sign WW (ss-WW) channel which has thus motivated their implementation in the {\tt POWHEG BOX} \cite{Chiesa:2019ulk}.
Such a hierarchy of the NLO contributions is not only due to the general characteristics of VBS but also to the large signal-over-background ratio in the ss-WW channel.
In other cases, like for vector-boson scattering into ZZ where the
background is large, such a hierarchy does not necessarily hold.
In this article we study these implications and give new insights in VBS and its irreducible background at the LHC.

The results presented in this article should be of great interest for experimental collaborations.
In particular, the ATLAS and CMS collaborations have already observed the EW $\PZ\PZ\Pj\Pj$ 
production \cite{Aad:2020zbq,Sirunyan:2017fvv,Sirunyan:2020alo}.
Along with state-of-the-art theoretical predictions, upcoming data will allow for deeper analyses of VBS at the LHC.

This article is structured as follows: In \refse{sec:calculation} the process under consideration is introduced.
In \refse{sec:results} the results are presented and analysed in detail.
Finally, \refse{sec:conclusion} contains a summary and concluding remarks.

\section{Description of the calculation}
\label{sec:calculation}

The present article complements \citere{Denner:2020zit} by providing the two missing NLO contributions of
orders $\order{\alphas^2 \alpha^5}$ and $\order{\alphas^3 \alpha^4}$.
Therefore, in the following, we often refer to this reference, and the interested reader is invited to look into it.
For the sake of simplicity some information is not repeated here as we focus on the salient features of the newly computed contributions.

\subsection{The process}
\label{sec:process}

\begin{figure}
\setlength{\parskip}{1ex}
\begin{subfigure}[c]{0.33\textwidth}
\centering
\captionsetup{skip=0pt}
\caption{}
\includegraphics[page=9,scale=0.9]{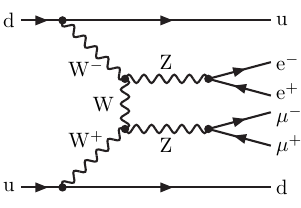}
\label{fig:born_qq_vbs}
\end{subfigure}
\begin{subfigure}{0.33\textwidth}
\centering
\captionsetup{skip=0pt}
\caption{}
\includegraphics[page=10,scale=0.9]{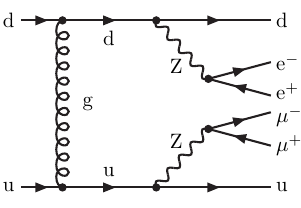}
\label{fig:born_QCD}
\end{subfigure}%
\begin{subfigure}[c]{0.33\textwidth}
\centering
\captionsetup{skip=0pt}
\caption{}
\includegraphics[page=11,scale=0.9]{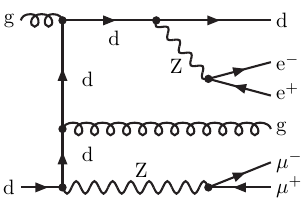}
\label{fig:born_gg}
\end{subfigure}%
\caption{Examples of tree-level Feynman diagrams: EW (left) and QCD
  (middle and right).}
\label{fig:lo}
\end{figure}

The physical process under investigation is 
\begin{equation}
\Pp\Pp \to \Pe^+\Pe^-\mu^+\mu^-\Pj\Pj+X 
\end{equation}
at the LHC.
In the same way as all processes containing VBS contributions, its
cross section possesses three leading-order (LO) components:
an EW one of order $\order{\alpha^6}$ including VBS,
an interference of order $\order{\alphas \alpha^5}$,
and a QCD contribution of order $\order{\alpha^2_{\rm s} \alpha^4}$.
Exemplary Feynman diagrams of order $\order{{g}^6}$ and
$\order{{g}^2_{\rm s} {g}^4}$ are shown in \reffi{fig:lo}. 
At order $\order{{g}^2_{\rm s} {g}^4}$, besides partonic processes
involving four quarks, also processes with two quarks and two gluons
appear (see \reffi{fig:born_gg}) with all possible crossings of gluons
and quarks.

At NLO, there are four orders contributing:
$\order{\alpha^7}$, $\order{\alphas \alpha^6}$,
$\order{\alpha^2_{\rm s} \alpha^5}$, and $\order{\alpha^3_{\rm s} \alpha^4}$.
The first two have been computed without approximations in \citere{Denner:2020zit}.
The latter has been evaluated in \citere{Campanario:2014ioa}, while the order $\order{\alpha^2_{\rm s} \alpha^5}$ is calculated here for the first time.
Given that the first two orders are discussed in detail in
\citere{Denner:2020zit}, 
the discussion in the following is focused on the two remaining
orders, which receive contributions from partonic processes involving
four quarks as well as from those with two quarks and gluons.

\begin{figure}
\begin{subfigure}[b]{0.33\textwidth}
\centering
\captionsetup{skip=0pt}\caption{}
\includegraphics[page=8,scale=0.9]{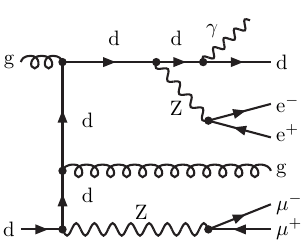}
\label{fig:real_gs2_g5}
\end{subfigure}
\begin{subfigure}[b]{0.33\textwidth}
\centering
\captionsetup{skip=0pt}\caption{}
\includegraphics[page=2,scale=0.9]{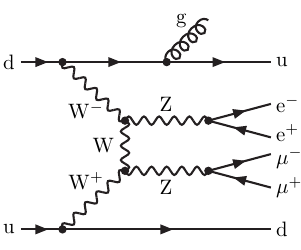}
\label{fig:real_gs_g6}
\end{subfigure}%
\begin{subfigure}[b]{0.33\textwidth}
\centering
\captionsetup{skip=0pt}\caption{}
\includegraphics[page=3,scale=0.9]{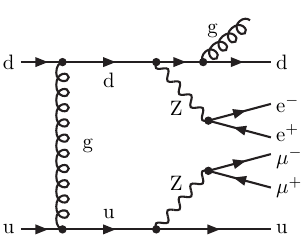}
\label{fig:real_gs3_g4}
\end{subfigure}%
\caption{Examples of real tree-level Feynman diagrams: photon emission
  (left) and gluon emission off LO EW (middle) and QCD diagrams (right).}
\label{fig:real_nlo}
\end{figure}

\begin{figure}
\begin{subfigure}[b]{0.33\textwidth}
\centering
\captionsetup{skip=0pt}\caption{}
\includegraphics[page=4,scale=0.9]{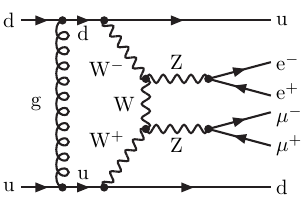}
\label{fig:loop_gs2_g6_qcd}
\end{subfigure}
\begin{subfigure}[b]{0.33\textwidth}
\centering
\captionsetup{skip=0pt}\caption{}
\includegraphics[page=7,scale=0.9]{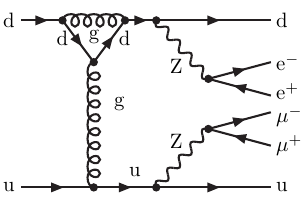}
\label{fig:loop_gs4_g4_vert}
\end{subfigure}%
\begin{subfigure}[b]{0.33\textwidth}
\centering
\captionsetup{skip=0pt}\caption{}
\includegraphics[page=6,scale=0.9]{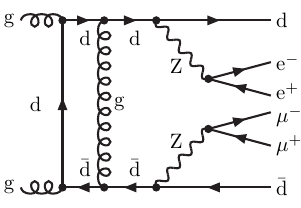}
\label{fig:loop_gs4_g4}
\end{subfigure}%
\caption{Examples of virtual one-loop Feynman diagrams: mixed EW--QCD
  loop (left) and purely QCD loops (middle and right).}
\label{fig:virt_nlo}
\end{figure}

The order $\order{\alpha^2_{\rm s} \alpha^5}$ is made of corrections of both QCD and EW types [as the order $\order{\alphas \alpha^6}$].
In particular, it features both types of real corrections: photon
emissions and emissions of QCD partons.
The first contribution entails squared matrix elements of photon-emission diagrams of order $\order{{g}^2_{\rm s} {g}^5}$ as shown in \reffi{fig:real_gs2_g5}.
The second one is made of interferences of QCD real matrix elements of orders $\order{{g}_{\rm s} {g}^6}$ (shown in \reffi{fig:real_gs_g6})
and $\order{{g}^3_{\rm s} {g}^4}$ (shown in \reffi{fig:real_gs3_g4}).
Their infrared (IR) singularities are compensated by the corresponding virtual corrections,
which also receive two types of contributions.
The first one consists of one-loop amplitudes of order $\order{{g}^2_{\rm s} {g}^6}$ (as shown in \reffi{fig:loop_gs2_g6_qcd}) interfered 
with tree-level amplitudes of order $\order{{g}^2_{\rm s} {g}^4}$ (\reffi{fig:born_QCD}).
The second type is furnished by one-loop amplitudes of order
$\order{{g}^4_{\rm s} {g}^4}$ (as shown in
\reffis{fig:loop_gs4_g4_vert}) interfered with tree-level
amplitudes of order $\order{{g}^6}$ (\reffi{fig:born_qq_vbs}).  We
note that the two types of NLO corrections cannot be defined in an
IR-finite way on the basis of Feynman diagrams in a full computation
\cite{Biedermann:2017bss}, \emph{i.e.}\ without relying on any
approximations \cite{Ballestrero:2018anz}. For instance, the diagram
in \reffi{fig:loop_gs2_g6_qcd} can be viewed as an EW correction to a
LO diagram of order $\order{{g}^2_{\rm s} {g}^4}$ or as a QCD
correction to a LO diagram of order $\order{g^6}$.

As explained above, at the order $\order{\alpha^2_{\rm s} \alpha^5}$,
some of the real corrections are made of a photon emission of a LO QCD amplitude.
It also means that in the final state one can have a photon as well as one (or two) gluon(s).
In the case where a hard photon is recombined with a soft gluon in a single jet, the QCD singularity associated to the soft gluon is not accounted for by the QED subtraction term.
To avoid such configurations, a veto is typically applied on jets that have a too large photon--jet energy fraction
\begin{equation}
 z_{\gamma} = \frac{E_\gamma}{E_\gamma+E_a} ,
\end{equation}
where $E_\gamma$ and $E_a$ are the energy of the photon and the QCD parton $a$, respectively.
In the present case, the numerical limit has been chosen to be
$z_{\gamma} < z_{\gamma, \rm cut}=0.7$.
By cutting into the collinear region, IR-safety is lost
  but can be restored upon including the nonperturbative fragmentation
  of quarks into photons via the fragmentation function \cite{Denner:2009gj,Denner:2010ia,Denner:2011vu,Denner:2014ina}.
Schematically, the NLO cross section in the relevant order can be written as
\begin{align}
\rd \sigma_{\rm NLO} = &\int_{n+1} \theta (z_{\gamma, \rm cut} - z_{\gamma})
\left[\rd \sigma_{\rm real} - \rd \sigma_{\rm dipole} \right] + \nonumber \\
&\int_{n} \left[\rd \sigma_{\rm virtual} + \int_1 \left( \rd \sigma_{\rm dipole} - \rd \sigma^{\gamma \; \rm coll}_{\rm dipole}(z_{\gamma, \rm cut}) \right) - \rd \sigma_{\rm frag}(z_{\gamma, \rm cut}) \right] ,
\end{align}
where the index at the integrals denotes the number of particles in
the corresponding phase spaces.
The modification of the dipoles due to the rejection of hard photons is thus encoded in
\begin{equation}
 \rd \sigma^{\gamma \; \rm coll}_{\rm frag}(z_{\gamma, \rm cut}) 
= \theta (z_{\gamma}- z_{\gamma, \rm cut}) \rd \sigma_{\rm dipole} .
\end{equation}
In addition, the contributions of final-state quarks fragmenting into a hard photon are accounted for by
\begin{align}
 \rd \sigma_{\rm frag}(z_{\gamma, \rm cut}) = \sum_i \rd \sigma_{\rm
   Born} \int^1_{z_\gamma, \rm cut} \rd z_\gamma \, D_{q_i\to\gamma} (z_\gamma) ,
\end{align}
where the sum runs over all the final state quarks and $D_{q_i\to\gamma}$ is the quark-to-photon fragmentation function.

For the present computation, the fit parameters entering the fragmentation function have been taken from \citere{Buskulic:1995au} and read
\begin{equation}
\mu_0 = 0.14 \GeV, \qquad C = -13.26 .
\end{equation}

Finally, the order $\order{\alpha^3_{\rm s} \alpha^4}$ is made of NLO QCD corrections only.
This means that it contains squares of amplitudes with a gluon attached to
the $\order{{g}^2_{\rm s} {g}^4}$ diagrams as shown in
\reffi{fig:real_gs3_g4}, furnishing the 
real corrections of order $\order{\alpha^3_{\rm s} \alpha^4}$.
The virtual contributions are made of one-loop amplitudes of order $\order{{g}^4_{\rm s} {g}^4}$ (as shown in \reffis{fig:loop_gs4_g4_vert} and \ref{fig:loop_gs4_g4}) 
interfered with tree-level amplitudes of order
$\order{{g}^2_{\rm s} {g}^4}$ (\reffi{fig:born_QCD}).

Note that we do not consider contributions with external bottom quarks in any of the LO or NLO contributions.
These are small or can be removed in experimental analyses with the help of bottom-jet vetoes \cite{Denner:2019tmn,Denner:2020zit}.

\subsection{Details and validation}
\label{sec:details}

The results presented here have been produced using the Monte Carlo program \mocanlo and the matrix-element generator \recola.
The program \mocanlo is able to compute arbitrary processes within the SM at NLO QCD and EW accuracy.
In order to obtain a fast integration for high-multiplicity processes with many resonances,
it takes advantage of phase-space mappings similar to the ones of
\citeres{Berends:1994pv,Denner:1999gp,Dittmaier:2002ap}. 
On the other hand, \recola~\cite{Actis:2012qn,Actis:2016mpe} is a general tree and one-loop matrix-element provider.
It relies on the \collier library~\cite{Denner:2014gla,Denner:2016kdg} that provides numerically the one-loop scalar
\cite{'tHooft:1978xw,Beenakker:1988jr,Dittmaier:2003bc,Denner:2010tr}
and tensor integrals \cite{Passarino:1978jh,Denner:2002ii,Denner:2005nn}.
This combination has already shown to work well for high-multiplicity processes ($2\to6$ and beyond).
Many of these applications concerned VBS processes \cite{Biedermann:2016yds,Biedermann:2017bss,Ballestrero:2018anz,Denner:2019tmn,Pellen:2019ywl,Denner:2020zit}.
During the course of these studies, the set of tools was verified against {\sc BONSAY+}\openloops \cite{Denner:2019tmn} for WZ scattering,
the Monte Carlo {\sc BBMC} for same-sign W scattering \cite{Biedermann:2016yds,Biedermann:2017bss}, 
and against {\sc Powheg} \cite{Chiesa:2019ulk} for same-sign W scattering at order $\order{\alpha^7}$.
For non-VBS processes, it was checked against {\sc Sherpa} in \citeres{Biedermann:2017yoi,Brauer:2020kfv} 
and a multitude of tools in \citere{Bendavid:2018nar} for di-boson production at NLO EW.

Finally, in \citere{Denner:2020zit}, the partonic processes $\Pu\Pd\to\Pe^{+}\Pe^{-}\mu^{+}\mu^{-}\Pu\Pd$,
$\Pu\Ps\to\Pe^{+}\Pe^{-}\mu^{+}\mu^{-}\Pd\Pc$,
$\Pu\Pu\to\Pe^{+}\Pe^{-}\mu^{+}\mu^{-}\Pu\Pu$, and
$\Pu\bar\Pc\to\Pe^{+}\Pe^{-}\mu^{+}\mu^{-}\Pd\bar\Ps$ at order
$\mathcal{O}{\left(\alpha^7\right)}$ were successfully compared between \mocanlorecola and {\sc BBMC+}\recola.
Also, selected contributions of order $\mathcal{O}{\left(\alphas\alpha^6\right)}$ were found in agreement within integration errors.
For the present article, about 10 representative partonic channels
were confirmed to agree between \mocanlo and {\sc BBMC} at the NLO orders
$\order{\alpha^2_{\rm s} \alpha^5}$ and $\order{\alpha^3_{\rm s}
  \alpha^4}$ within statistical errors at the level of a few per cent.

In \mocanlo and {\sc BBMC}, the subtraction of IR divergences in the
real radiation is realised with the help of the Catani--Seymour dipole formalism \cite{Catani:1996vz,Dittmaier:1999mb,Dittmaier:2008md}.
Within this method, it is possible to vary the
$\alpha_{\mathrm{dipole}}$ parameter~\cite{Nagy:1998bb}, which restricts
the subtraction to IR-singular regions of the phase space.
Obtaining consistent results with different values $\alpha_{\mathrm{dipole}}$ ensures the correctness of the subtraction mechanism.
We have therefore performed two full computations of all NLO orders using \mocanlorecola with $\alpha_{\mathrm{dipole}}=1$ and $\alpha_{\mathrm{dipole}}=10^{-2}$.
Full statistical agreement has been found, and the results presented here are the ones obtained with $\alpha_{\mathrm{dipole}}=10^{-2}$.
Also, for the computation of order $\order{\alpha^2_{\rm s}
  \alpha^5}$ two different modes of {\sc Collier} have been
successfully compared.
The results presented here have been obtained with the {\sc COLI} mode throughout.
In all computations, the complex-mass scheme \cite{Denner:1999gp,Denner:2005fg,Denner:2006ic,Denner:2019vbn} is utilised.

Finally, the implementation of the fragmentation function closely follows \citeres{Denner:2009gj,Denner:2010ia,Denner:2011vu,Denner:2014ina}.
Its implementation in \mocanlo has been validated against {\sc BBMC} which was used in \citere{Denner:2014ina} 
for the computation of the EW corrections to $\Pp\Pp\to\ell^+\ell^-\Pj\Pj+X$.
Also, this implementation was employed in \citere{Brauer:2020kfv} for $\Pp\Pp\to\Pe^+\nu_\Pe\mu^-\bar\nu_\mu\Pj+X$ 
where it was compared against an approximate treatment.

\section{Numerical results}
\label{sec:results}

\subsection{Input parameters and event selection}

The input parameters and event selection used in the present article
are exactly the same as those in \citere{Denner:2020zit}.
For completeness we reproduce them here.

\subsubsection*{Input parameters}
\label{ssec:InputParameters}

The calculation is done for LHC characteristics and a centre-of-mass energy of $13\TeV$. 
The set of parton distribution functions (PDF) NLO NNPDF-3.1 Lux QED with $\alphas(\MZ) = 0.118$ \cite{Ball:2014uwa,Bertone:2017bme} and with fixed $N_\text{F}=5$ flavour scheme is used throughout.
All collinear initial-state splittings are treated by the ${\overline{\rm MS}}$ redefinition of the PDF.
The PDF set is interfaced to our Monte Carlo programs using LHAPDF~\cite{Andersen:2014efa,Buckley:2014ana}.

The renormalisation and factorisation scales, $\mu_{\rm ren}$ and
$\mu_{\rm fac}$, are chosen to be
\begin{equation}
\label{eq:defscale}
\mu_0 = \sqrt{p_{\rT,\Pj_1}\, p_{\rT, \Pj_2}}
\end{equation}
for all contributions, where $\Pj_1$ and $\Pj_2$ are the two hardest (in $p_{\rm T}$)
identified jets (tagging jets).
This is the scale choice that is usually used for VBS processes in the
literature and that has been also employed in \citere{Denner:2020zit}.
In cross sections and differential distributions, the scale uncertainty
is obtained with the standard 7-point scale variation, \ie 
observables are evaluated for the 7~pairs of renormalisation and factorisation scales
\begin{equation}
(\mu_\mathrm{ren}/\mu_0,\mu_\mathrm{fact}/\mu_0) = (0.5,0.5),
(0.5,1),(1,0.5),(1,1),(1,2),(2,1),(2,2),
\end{equation}
and the scale variation is determined from the resulting envelope.

The electromagnetic coupling is obtained through the $G_\mu$ scheme \cite{Denner:2000bj} via
\begin{equation}
  \alpha = \frac{\sqrt{2}}{\pi} G_\mu \MW^2 \left( 1 -
    \frac{\MW^2}{\MZ^2}\right)  \qquad \text{with}  \qquad   {\GF    = 1.16638\times 10^{-5}\GeV^{-2}},
\end{equation}
where $\GF$ is the Fermi constant.
The numerical values of the masses and widths used as input in the
numerical simulation read \cite{ParticleDataGroup:2020ssz}
%
\begin{alignat}{2}
\label{eqn:ParticleMassesAndWidths}
\Mt   &=  173.0\GeV,      & \quad \quad \quad \Gt &= 0 \GeV,  \nonumber \\
\MZOS &=  91.1876\GeV,      & \quad \quad \quad \GZOS &= 2.4952\GeV,  \nonumber \\
\MWOS &=  80.379\GeV,       & \GWOS &= 2.085\GeV,  \nonumber \\
M_{\rm H} &=  125.0\GeV,    &  \GH  &=  4.07 \times 10^{-3}\GeV.
\end{alignat}
Everywhere, it is assumed that $\Mb =0\GeV$ and as mentioned above no partonic channels with external bottom quarks are considered.
For the massive gauge bosons (W and Z), the pole masses and widths utilised in the calculation are obtained from the on-shell (OS) values \cite{Bardin:1988xt} using
\begin{equation}
M_V = \frac{\MVOS}{\sqrt{1+(\GVOS/\MVOS)^2}}\,,\qquad  
\Gamma_V = \frac{\GVOS}{\sqrt{1+(\GVOS/\MVOS)^2}}.
\end{equation}

\subsubsection*{Event selection}
\label{se:eventselection}

The event selection of the present computation is largely inspired from the CMS analyses \cite{Sirunyan:2017fvv,Sirunyan:2020alo}.
The process features four charged leptons and two jets in the final state.
Quarks and gluons with pseudorapidity $|\eta|<5$ are clustered into jets with the anti-$k_{\rT}$ algorithm \cite{Cacciari:2008gp} using $R=0.4$.
Following the same method and radius parameter, the photons are recombined with the final-state jets or leptons. 

The four leptons $\ell$ must fulfil
\begin{align}
\ptsub{\Pl} >  20\GeV,\qquad |\eta_{\Pl}| < 2.5, \qquad 
\Delta R_{\Pl\Pl'} > 0.05, \qquad M_{\Pl^+\Pl^{\prime-}} > 4 \GeV,
\end{align}
where $\Pl$ and $\Pl'$ are any leptons while $\Pl^+$ and $\Pl^{\prime-}$ are oppositely charged leptons regardless of their flavour.
On top of these cuts, the invariant masses of the leptonic decay
products of the Z bosons are restricted to
\begin{align}
\label{eq:mz}
60 \GeV < M_{\Pl^+\Pl^-} < 120 \GeV,\qquad \Pl=\Pe,\mu.
\end{align}

Once the jet clustering is performed, at least two jets still have to fulfil the criteria
\begin{align}
\label{eq:jet1}
\ptsub{\Pj} >  30\GeV, \qquad |\eta_\Pj| < 4.7,\qquad\Delta R_{\Pj\Pl} > 0.4 .
\end{align}
Such jets are then denoted as \emph{identified} jets.
Out of these, the two with the highest transverse momenta are called
{\em tagging jets.}

In our default setup (\emph{inclusive setup} for short) the tagging jets have to fulfil the constraint
\begin{align}
\label{eq:vbscuts}
M_{\Pj_1 \Pj_2} > 100\GeV .
\end{align}
We consider in addition a setup (\emph{VBS setup}) where the last cut is
replaced by a stronger one
\begin{align}
\label{eq:vbscuts2}
M_{\Pj_1 \Pj_2} > 500\GeV .
\end{align}

\subsection{Cross sections}

In this section, various cross sections at LO and NLO accuracy are discussed.
We start by recalling the LO cross sections computed in \citere{Denner:2020zit} in \refta{tab:LO}.
\begin{table}
\sisetup{group-digits=false}
\centering
\begin{tabular}{ccccc|c}
\toprule
Order
    & $\order{           \alpha^6 }$
    & $\order{ \alphas   \alpha^5 }$
    & $\order{ \alphas^2   \alpha^4 }$
    & Sum
    & $\order{ \alphas^4   \alpha^4 }$ \\
\midrule
\multicolumn{2}{l}{$M_{\Pj_1\Pj_2}> 100\GeV$} \\
\midrule
$\sigma_{\mathrm{LO}}^{4\Pq} [\si{\femto\barn}]$
    & \num{ 0.097683 +- 0.000002}
    & \num{ 0.008628 +- 0.000001}
    & \num{ 0.22138 +- 0.00001}
    & \num{ 0.32770 +- 0.00001}
    & -
    \\
$\sigma_{\mathrm{LO}}^{2\Pq2\Pg} [\si{\femto\barn}]$
    & -
    & -
    & \num{ 0.84122 +- 0.00005}
    & \num{ 0.84122 +- 0.00005}
    & \num{ 0.1210 +- 0.0006}
    \\
$\sigma_{\mathrm{LO}} [\si{\femto\barn}]$
    & \num{ 0.097683 +- 0.000002}
    & \num{ 0.008628 +- 0.000001}
    & \num{ 1.06260  +- 0.00005}
    & \num{ 1.16891  +- 0.00005}
    & \num{ 0.1210 +- 0.0006}
    \\
$\textrm{fraction} [\si{\percent}]$
    & \num{8.36}
    & \num{0.74}
    & \num{90.91}
    & \num{100}
    & \num{10.35} \\
\midrule
\multicolumn{2}{l}{$M_{\Pj_1\Pj_2}> 500\GeV$} \\
\midrule
$\sigma_{\mathrm{LO}}^{4\Pq} [\si{\femto\barn}]$
    & \num{ 0.073676  +- 0.000003}
    & \num{ 0.005567  +- 0.000001}
    & \num{ 0.04623   +- 0.000004}
    & \num{ 0.12547   +- 0.000004}
    & -
    \\
$\sigma_{\mathrm{LO}}^{2\Pq2\Pg} [\si{\femto\barn}]$
    & -
    & -
    & \num{ 0.08992   +- 0.00002 }
    & \num{ 0.08992  +- 0.00002 }
    & \num{ 0.0135  +- 0.0003}
    \\
$\sigma_{\mathrm{LO}} [\si{\femto\barn}]$
    & \num{ 0.073676  +- 0.000003}
    & \num{ 0.005567  +- 0.000001}
    & \num{ 0.13614   +- 0.00002 }
    & \num{ 0.21539  +- 0.00002 }
    & \num{ 0.0135  +- 0.0003}
    \\
$\textrm{fraction} [\si{\percent}]$
    & \num{34.21}
    & \num{2.58}
    & \num{63.21}
    & \num{100}
    & \num{6.24} \\
\bottomrule
\end{tabular}
\caption{LO cross section of the individual
orders $\order{\alpha^6}$, $\order{\alphas\alpha^5}$,
$\order{\alphas^2\alpha^4}$, and their sum for $\Pp\Pp \to
\Pe^+\Pe^-\mu^+\mu^-\Pj\Pj+X$ at $13\TeV$ CM energy. 
Contributions of partonic channels with four external quarks, 
 $\sigma_{\mathrm{LO}}^{4\Pq}$, and those with two external quarks and
 two external gluons,
 $\sigma_{\mathrm{LO}}^{2\Pq2\Pg}$, are shown separately as well.
The loop-induced contribution of $\order{ \alphas^4 \alpha^4 }$ is 
also listed but does not enter the sum of contributions.
Each contribution is given in $\fb$ and as fraction relative to the
sum of the three contributions (in per cent).
While the numbers in the upper part of the table are for the inclusive setup, those in the lower part are for the VBS setup.
The digits in parentheses indicate integration errors.
}
\label{tab:LO}
\end{table}
As opposed to \citere{Denner:2020zit}, here the sum comprises only
the contributions of orders $\order{\alpha^6}$,
$\order{\alphas\alpha^5}$, and $\order{\alphas^2\alpha^4}$, while 
the loop-induced contribution of order $\order{ \alphas^4 \alpha^4 }$ 
from $\Pg\Pg\to\Pe^+\Pe^-\mu^+\mu^-\Pg\Pg$
does not enter the sum and is only shown for completeness.
Contributions of partonic channels with four external quarks,
$\sigma_{\mathrm{LO}}^{4\Pq}$, and of those involving gluons,
 $\sigma_{\mathrm{LO}}^{2\Pq2\Pg}$, which enter at order
 $\order{\alphas^2\alpha^4}$,  are shown separately as well.
When referring to relative corrections in the following, they are
always normalised to the above LO sum, and the  loop-induced
contribution is not included in the NLO predictions.
\refta{tab:LO} clearly shows that the QCD contribution of order
$\order{\alphas^2\alpha^4}$ is dominating and reaches $91\%$ of
the LO sum for $M_{\Pj_1\Pj_2}> 100\GeV$.
When applying the tighter VBS event selection, $M_{\Pj_1\Pj_2}>
500\GeV$, it is reduced to $63\%$, demonstrating hence the impact of
such cuts in enhancing the EW contribution.
While the partonic channels involving quarks and gluons contribute
$79\%$ at $\order{\alphas^2\alpha^4}$ and $72\%$ of the LO sum for the
inclusive case, their share is diminished to $66\%$ and $42\%$,
respectively, for the VBS setup.
The tighter VBS cuts also reduce the relative contribution of the
loop-induced process from $10\%$ down to $6\%$. 
Including 7-point scale variations, the full LO cross section
comprising the orders $\order{\alpha^6}$, $\order{\alpha_s \alpha^5}$,
and $\order{\alphas^2\alpha^4}$ reads
\begin{align}
 \sigma_{\textrm{LO}} ={} &  1.16879(5)^{+29.1\%}_{-20.7\%}
\qquad \text{ for } \quad M_{\Pj_1\Pj_2}> 100\GeV,\notag\\
 \sigma_{\textrm{LO}} ={} &0.21539(2)^{+26.0\%}_{-18.3\%} 
\qquad \text{ for } \quad M_{\Pj_1\Pj_2}> 500\GeV.
\end{align}
With $30\%$ the LO scale dependence has the typical order of magnitude
for a cross section of order $\order{\alphas^2}$. It is somewhat
reduced for $M_{\Pj_1\Pj_2}> 500\GeV$ owing to the larger fraction of
the EW contribution.

The full NLO cross section, including orders
$\mathcal{O}{\left(\alpha^{7}\right)}$,
$\mathcal{O}{\left(\alphas\alpha^{6}\right)}$,
$\mathcal{O}{\left(\alphas^{2}\alpha^{5}\right)}$, and
$\mathcal{O}{\left(\alphas^{3}\alpha^{4}\right)}$, reads with
7-point scale variations
\begin{align}
 \sigma_{\textrm{NLO}} ={} &  1.040(2)^{+0.99\%}_{-9.27\%}
\qquad \text{ for } \quad M_{\Pj_1\Pj_2}> 100\GeV,\notag\\
 \sigma_{\textrm{NLO}} ={} & 0.194(1)^{+0.64\%}_{-5.40\%} 
\qquad \text{ for } \quad M_{\Pj_1\Pj_2}> 500\GeV,
\end{align}
\ie the scale dependence is reduced by more than a factor 3 when including
NLO corrections.

In \refta{tab:NLO} the NLO corrections of orders 
$\mathcal{O}{\left(\alpha^{7}\right)}$,
$\mathcal{O}{\left(\alphas\alpha^{6}\right)}$,
$\mathcal{O}{\left(\alphas^{2}\alpha^{5}\right)}$, and
$\mathcal{O}{\left(\alphas^{3}\alpha^{4}\right)}$ are shown
separately, split into contributions from four-quark processes and
gluon--quark processes where appropriate.
\begin{table}
\sisetup{group-digits=false}
\centering
\begin{tabular}{cccccc}
\toprule
Order
    & $\order{           \alpha^7 }$
    & $\order{ \alphas   \alpha^6 }$
    & $\order{ \alphas^2   \alpha^5 }$
    & $\order{ \alphas^3   \alpha^4 }$
    & Sum \\
\midrule
\multicolumn{2}{l}{$M_{\Pj_1\Pj_2}> 100\GeV$} \\
\midrule
$\Delta\sigma_{\mathrm{NLO}}^{4\Pq} [\si{\femto\barn}]$
    & \num{-0.01557 +- 0.00004}
    & \num{ 0.0231 +- 0.0001}
    & \num{-0.0162 +- 0.0001}
    & \num{-0.0444 +- 0.0007}
    & \num{-0.0531 +- 0.0007}
\\
$\Delta\sigma_{\mathrm{NLO}}^{2\Pq2\Pg} [\si{\femto\barn}]$
    & -
    & -
    & \num{-0.0673 +- 0.0001}
    & \num{-0.0081 +- 0.0018}
    & \num{-0.0754 +- 0.0018}
\\
$\Delta\sigma_{\mathrm{NLO}} [\si{\femto\barn}]$
    & \num{-0.01557 +- 0.00004}
    & \num{ 0.0231 +- 0.0001}
    & \num{-0.0835 +- 0.0001}
    & \num{-0.0525 +- 0.0019}
    & \num{-0.1285 +- 0.0019}
    \\
$\Delta\sigma_{\mathrm{NLO}}/\sigma_{\mathrm{LO}} [\si{\percent}]$
    & \num{-1.33 +- 0.01 }
    & \num{1.98 +- 0.01 }
    & \num{-7.14 +- 0.01}
    & \num{-4.49 +- 0.16 }
    & \num{-10.99 +- 0.16 } \\
\midrule
\multicolumn{2}{l}{$M_{\Pj_1\Pj_2}> 500\GeV$} \\
\midrule
$\Delta\sigma_{\mathrm{NLO}}^{4\Pq} [\si{\femto\barn}]$
    & \num{-0.01299  +- 0.00005}
    & \num{ 0.00008  +- 0.00025}
    & \num{-0.00476  +- 0.00009}
    & \num{ 0.0016  +- 0.0003}
    & \num{-0.0160  +- 0.0004}
\\
$\Delta\sigma_{\mathrm{NLO}}^{2\Pq2\Pg} [\si{\femto\barn}]$
    & -
    & -
    & \num{-0.00926  +- 0.00003}
    & \num{ 0.0041  +- 0.0006}
    & \num{-0.0051  +- 0.0007}
\\
$\Delta\sigma_{\mathrm{NLO}} [\si{\femto\barn}]$
    & \num{-0.01299  +- 0.00005}
    & \num{ 0.00008  +- 0.00025}
    & \num{-0.01402  +- 0.00009}
    & \num{ 0.0058  +- 0.0007}
    & \num{-0.0211  +- 0.0008}
    \\
$\Delta\sigma_{\mathrm{NLO}}/\sigma_{\mathrm{LO}} [\si{\percent}]$
    & \num{-6.03 +-0.02}
    & \num{+0.04 +-0.11}
    & \num{-6.51 +-0.04}
    & \num{+2.69 +-0.34}
    & \num{-9.81 +-0.36} \\
\bottomrule
\end{tabular}
\caption{NLO corrections for the process $\Pp\Pp \to \Pe^+\Pe^-\mu^+\mu^-\Pj\Pj+X$ at the orders 
$\mathcal{O}{\left(\alpha^{7}\right)}$, $\mathcal{O}{\left(\alphas\alpha^{6}\right)}$, $\mathcal{O}{\left(\alphas^{2}\alpha^{5}\right)}$, $\mathcal{O}{\left(\alphas^{3}\alpha^{4}\right)}$, 
and for the sum of the four NLO corrections.
Contributions of partonic channels involving four quarks, 
 $\sigma_{\mathrm{LO}}^{4\Pq}$, and those involving two quarks,
 $\sigma_{\mathrm{LO}}^{2\Pq2\Pg}$, are shown separately.
The contribution $\Delta\sigma_{\mathrm{NLO}}$ corresponds to the absolute correction for the central scale choice, while $\Delta \sigma_{\mathrm{NLO}}/\sigma_{\rm LO}$ gives the relative correction normalised to the sum of all LO contributions at the central scale.
The absolute contributions are expressed in femtobarn while the relative ones are expressed in per cent.
The statistical uncertainty from the Monte Carlo integration on the last digit is given in parentheses.}
\label{tab:NLO}
\end{table}
The relative corrections are normalised to the sum of the LO contributions of orders $\order{\alpha^6}$, $\order{\alphas\alpha^5}$, and $\order{\alphas^2\alpha^4}$. 
As for \refta{tab:LO}, two setups with $M_{\Pj_1\Pj_2}> 100\GeV$ and $M_{\Pj_1\Pj_2}> 500\GeV$ are displayed.
For the case $M_{\Pj_1\Pj_2}> 100\GeV$, the largest correction is the
one of order $\mathcal{O}{\left(\alphas^{2}\alpha^{5}\right)}$ with
$-7.1\%$, dominated by the EW corrections to the LO QCD-induced contribution,
followed by the one of order
$\mathcal{O}{\left(\alphas^{3}\alpha^{4}\right)}$ with $-4.5\%$, the
corresponding QCD corrections.
The corrections of order $\mathcal{O}{\left(\alpha^{7}\right)}$ and $\mathcal{O}{\left(\alphas\alpha^{6}\right)}$ are on the other hand between $1\%$ and $2\%$.
The picture changes when imposing the restriction $M_{\Pj_1\Pj_2}> 500\GeV$.
Then, the two largest NLO contributions are the EW corrections of orders
$\mathcal{O}{\left(\alpha^{7}\right)}$ and
$\mathcal{O}{\left(\alphas^{2}\alpha^{5}\right)}$ being both negative
and  between $-6\%$ and $-6.5\%$. The contributions of order
$\mathcal{O}{\left(\alphas^{2}\alpha^{5}\right)}$  are dominated by
partonic channels involving two quarks, which amount to $80\%$ for the
inclusive setup and $66\%$ for the VBS setup. At
$\mathcal{O}{\left(\alphas^{3}\alpha^{4}\right)}$, 
the four-quark channels dominate for $M_{\Pj_1\Pj_2}> 100\GeV$, while the gluon--quark channels
take over for $M_{\Pj_1\Pj_2}> 500\GeV$. In the sum of all NLO corrections 
contributions from four-quark and gluon--quark channels are of similar
size.

The hierarchy of NLO contributions is quite different from the one in
$\Pp\Pp\to\mu^+\nu_\mu\Pe^+\nu_{\Pe}\Pj\Pj+X$, which was computed in
\citere{Biedermann:2017bss}. 
There, the largest NLO corrections were those of order
$\mathcal{O}{\left(\alpha^{7}\right)}$ with $-13.2\%$ followed by
those of order $\mathcal{O}{\left(\alphas\alpha^{6}\right)}$ with
$-3.5\%$, while
the contributions of orders $\mathcal{O}{\left(\alphas^{2}\alpha^{5}\right)}$ and $\mathcal{O}{\left(\alphas^{3}\alpha^{4}\right)}$ were subleading and below $1\%$.
This different hierarchy between the NLO corrections for the two
signatures is linked to the rather different LO hierarchy between the
EW and QCD components, which originates to a considerable extent from
the appearance of partonic channels involving
gluons, which are not present for ss-WW. 
While the EW component of order $\mathcal{O}{\left(\alpha^{6}\right)}$
represents $86.5\%$ of the total LO in the case of ss-WW, it
contributes only $34.2\%$ in the case of ZZ for the most exclusive
setup. 
The LO hierarchy strongly determines the one of the NLO corrections.
For example, when normalising the corrections of order
$\mathcal{O}{\left(\alpha^{7}\right)}$ for VBS into ZZ
to the LO EW contribution of order $\mathcal{O}{\left(\alpha^{6}\right)}$,
one obtains $-17.6\%$ in line with the results of \citeres{Biedermann:2016yds,Biedermann:2017bss,Denner:2019tmn}.

Interestingly, the corrections of order
$\mathcal{O}{\left(\alphas^{2}\alpha^{5}\right)}$ are sizeable and
when normalised to the order
$\mathcal{O}{\left(\alphas^{2}\alpha^{4}\right)}$ they are about
$-8\%$ and $-10\%$ for $M_{\Pj_1\Pj_2}> 100\GeV$ and $M_{\Pj_1\Pj_2}>
500\GeV$, respectively.  Such a characteristics has not been observed
in the case of ss-WW where the
$\mathcal{O}{\left(\alphas^{2}\alpha^{5}\right)}$ corrections, with
the same normalisation, are only of the order of $0.2\%$. The reason
of this different behaviour is revealed when considering the separate
contributions to the $\mathcal{O}{\left(\alphas^{2}\alpha^{5}\right)}$
corrections, namely EW corrections to the LO
$\mathcal{O}{\left(\alphas^{2}\alpha^{4}\right)}$ contributions and
QCD corrections to the LO
$\mathcal{O}{\left(\alphas\alpha^{5}\right)}$ interferences. While
this split is obvious for real radiation contributions, it cannot be
unambiguously done for virtual corrections \cite{Biedermann:2017bss}.
For the sake of this analysis, we count all contributions involving EW
bosons in the loop (see \reffi{fig:loop_gs2_g6_qcd}) as EW
corrections, while QCD corrections include only diagrams with merely
gluons and quarks in the loop (see \reffis{fig:loop_gs4_g4_vert} and
\ref{fig:loop_gs4_g4}). For ss-WW the EW
$\mathcal{O}{\left(\alphas^{2}\alpha^{5}\right)}$ corrections in fact
amount to $-12.3\%$ of the LO
$\mathcal{O}{\left(\alphas^{2}\alpha^{4}\right)}$ contributions. These
corrections are, however, almost completely cancelled by the
$\mathcal{O}{\left(\alphas^{2}\alpha^{5}\right)}$ corrections of QCD
origin resulting in a net correction below a per cent. This
cancellation is possible since the LO interference of order
$\mathcal{O}{\left(\alphas\alpha^{5}\right)}$ reaches $30\%$ of the LO
QCD contribution. For VBS into ZZ, on the other hand, the LO
interference is only $4\%$ and $0.8\%$ of the LO QCD contribution for
$M_{\Pj_1 \Pj_2} > 100\GeV$ and $M_{\Pj_1 \Pj_2} > 500\GeV$,
respectively. As a result, the NLO QCD corrections of order
$\mathcal{O}{\left(\alphas^2\alpha^{5}\right)}$ are small and reduce
the NLO EW corrections to the LO
$\mathcal{O}{\left(\alphas^2\alpha^{4}\right)}$ only slightly from 
$-8.2\%$ to $-7.9\%$ for $M_{\Pj_1 \Pj_2} > 100\GeV$ and from $-11.6\%$ to
$-10.3\%$ or $M_{\Pj_1 \Pj_2} > 500\GeV$.
The situation is different for some partonic channels (see below).
Note also that for ZZ the behaviour of the $\mathcal{O}{\left(\alphas\alpha^{5}\right)}$
corrections is largely determined by the partonic channels involving
gluons, which are absent for ss-WW.

Some qualitative understanding of the NLO EW corrections of order
$\mathcal{O}{\left(\alphas^{2}\alpha^{5}\right)}$ can be obtained from
an high-energy approximation of the virtual corrections in the pole
approximation, \ie for the process
$\Pp\Pp\to\PZ\PZ\Pj\Pj\to\Pe^+\Pe^-\mu^+\mu^-\Pj\Pj+X$. To keep it
simple, we restrict the discussion to the dominant contributions of
left-handed quarks and transverse Z bosons and consider only the
double EW logarithms, the collinear single EW logarithms, and the
single EW logarithms resulting from parameter renormalisation. We do
not include the angular-dependent leading logarithms, which have a
much more complicated structure. Based on \citere{Denner:2000jv}, we
find for the EW correction factor to the cross section of order
$\mathcal{O}{\left(\alphas^{2}\alpha^{4}\right)}$ for
$\Pq\Pq\to\PZ\PZ\Pq\Pq$ ($n_\Pq=2$) and $\Pq\Pg\to\PZ\PZ\Pq\Pg$
($n_\Pq=1$) in the leading-logarithmic approximation:
\begin{align}\label{eq:LLcorr}
\delta_{\mathrm{LL}} {}={}& 
\frac{\alpha}{4\pi} \left\{-2 C^{\mathrm{EW}}_W
F_{\Pq}
-2 n_\Pq C^{\mathrm{EW}}_{\Pq}\right\}
\ln^2\left(\frac{Q^2}{\MW^2}\right)
+\frac{\alpha}{4\pi} 
6 n_\Pq C^{\mathrm{EW}}_{\Pq}
\ln\left(\frac{Q^2}{\MW^2}\right).
\end{align}
The constants are given by 
\begin{align}
 C^{\mathrm{EW}}_{\PW} ={}& \frac{2}{\sw^2}\approx 8.97,\qquad
 C^{\mathrm{EW}}_{\Pq} = \frac{1+26\cw^2}{36\cw^2\sw^2}\approx
 3.40,
\end{align}
and
\begin{align}
F_{\Pq}={}&\frac{I^3_{\mathrm{w},\Pq}(1-\sw^2)}{I^3_{\mathrm{w},\Pq}-\sw^2
  Q_\Pq},\qquad
F_{\Pu}\approx1.106, \qquad F_{\Pd}\approx0.913,
\end{align}
where $I^3_{\mathrm{w},\Pq}$ and $Q_\Pq$ denote the weak isospin and
relative charge of the quark $\Pq$, respectively, while $\sw$ and
$\cw$ stand for the sine and cosine of the EW mixing angle.  The
correction factor \refeq{eq:LLcorr} also applies unchanged to
processes resulting from crossing symmetry and/or if one of the quark
pairs is replaced by a different quark pair with the same quantum numbers,
\eg $\Pu\to\Pc$. For processes involving quarks with different quantum
numbers, the simple formula \refeq{eq:LLcorr} does not hold, however,
in the limit of vanishing hypercharge it is valid for arbitrary quark
combinations with $F_{\Pq}=1$. Since the correction factor
\refeq{eq:LLcorr} only describes EW corrections to the LO cross section of order
$\mathcal{O}{\left(\alphas^{2}\alpha^{4}\right)}$, this approximation
does not apply if QCD corrections to the LO interference of order
$\mathcal{O}{\left(\alphas\alpha^{5}\right)}$ are sizeable, which is
usually the case if the LO EW cross section  of order
$\mathcal{O}{\left(\alpha^{6}\right)}$ is comparable or larger than
the QCD one  of order
$\mathcal{O}{\left(\alphas^{2}\alpha^{4}\right)}$.
This happens, in particular, for the partonic processes
$\Pd\bar\Pd\to\Pe^+\Pe^-\mu^+\mu^-\Pu\bar\Pu$, 
$\Pu\bar\Pu\to\Pe^+\Pe^-\mu^+\mu^-\Pd\bar\Pd$, 
$\Pd\Pu\to\Pe^+\Pe^-\mu^+\mu^-\Pd\Pu$, and
$\bar\Pd\bar\Pu\to\Pe^+\Pe^-\mu^+\mu^-\bar\Pd\bar\Pu$, where the
contributions of orders $\mathcal{O}{\left(\alphas^{2}\alpha^{4}\right)}$
and $\mathcal{O}{\left(\alpha^{6}\right)}$ are comparable.

The only free parameter of \refeq{eq:LLcorr} is the scale $Q$.  While
for a $2\to2$ process one usually picks the centre-of-mass energy, a
$2\to4$ process involves many more scales. As a typical scale, we
choose the invariant mass of the $\PZ\PZ$ pair,
$Q=M_{\PZ\PZ}=M_{4\ell}$, which is the scale appearing in the leading
double logarithms multiplied with the largest coupling factor.
Determining the average of $M_{4\ell}$ from LO distributions, yields
values for $Q$ in the range $330$--$470\GeV$ for all types of
processes in the inclusive setup. Using $Q=M_{4\ell}$ in
\refeq{eq:LLcorr} event by event results in correction factors that
agree with the full calculation within about $2\%$ for all partonic
processes with external gluons, which dominate the cross section at
$\mathcal{O}{\left(\alphas^{2}\alpha^{4}\right)}$, and within $3\%$
for most four-quark processes.  In the VBS setup, we find average
values for $Q$ within $390$--$550\GeV$ for four-quark processes and
$370$--$530\GeV$ for gluon--quark processes  as well as correction
factors $\delta_{\mathrm{LL}}$ that agree with the full relative
corrections within about $3\%$ for the gluonic processes and within
$4\%$ for most four-quark processes.  The approximation fails completely
for the four-quark processes with sizeable
$\mathcal{O}{\left(\alpha^{6}\right)}$ contributions mentioned at the
end of the preceding paragraph.
Besides those, the approximation is worse than the above-stated figures only for
$\Pd\bar\Pd\to\Pe^+\Pe^-\mu^+\mu^-\Pc\bar\Pc$ and
$\Pu\bar\Pu\to\Pe^+\Pe^-\mu^+\mu^-\Ps\bar\Ps$,
owing to somewhat larger QCD corrections to
the EW LO interference, as well as for
$\Pu\Pu\to\Pe^+\Pe^-\mu^+\mu^-\Pu\Pu$, which has the highest scales
$M_{4\ell}$ among all partonic processes.  In general, the logarithmic
approximation predicts larger corrections for processes involving
up-type quarks as compared to those with down-type quarks, which is
also observed in the full calculation. In summary, the simple
approximation \refeq{eq:LLcorr} with the scale choice $Q=M_{4\ell}$
reproduces the $\mathcal{O}{\left(\alphas^{2}\alpha^{5}\right)}$
corrections to the fiducial cross section within a few per cent for
all partonic processes that are dominated by the LO
$\mathcal{O}{\left(\alphas^{2}\alpha^{4}\right)}$ contributions and
for the sum of all partonic channels.

\subsection{Differential distributions}

\begin{figure}
\setlength{\parskip}{-4ex}
\begin{subfigure}{0.49\textwidth}
\centering
\subcaption{}
\includegraphics[width=1.\linewidth]{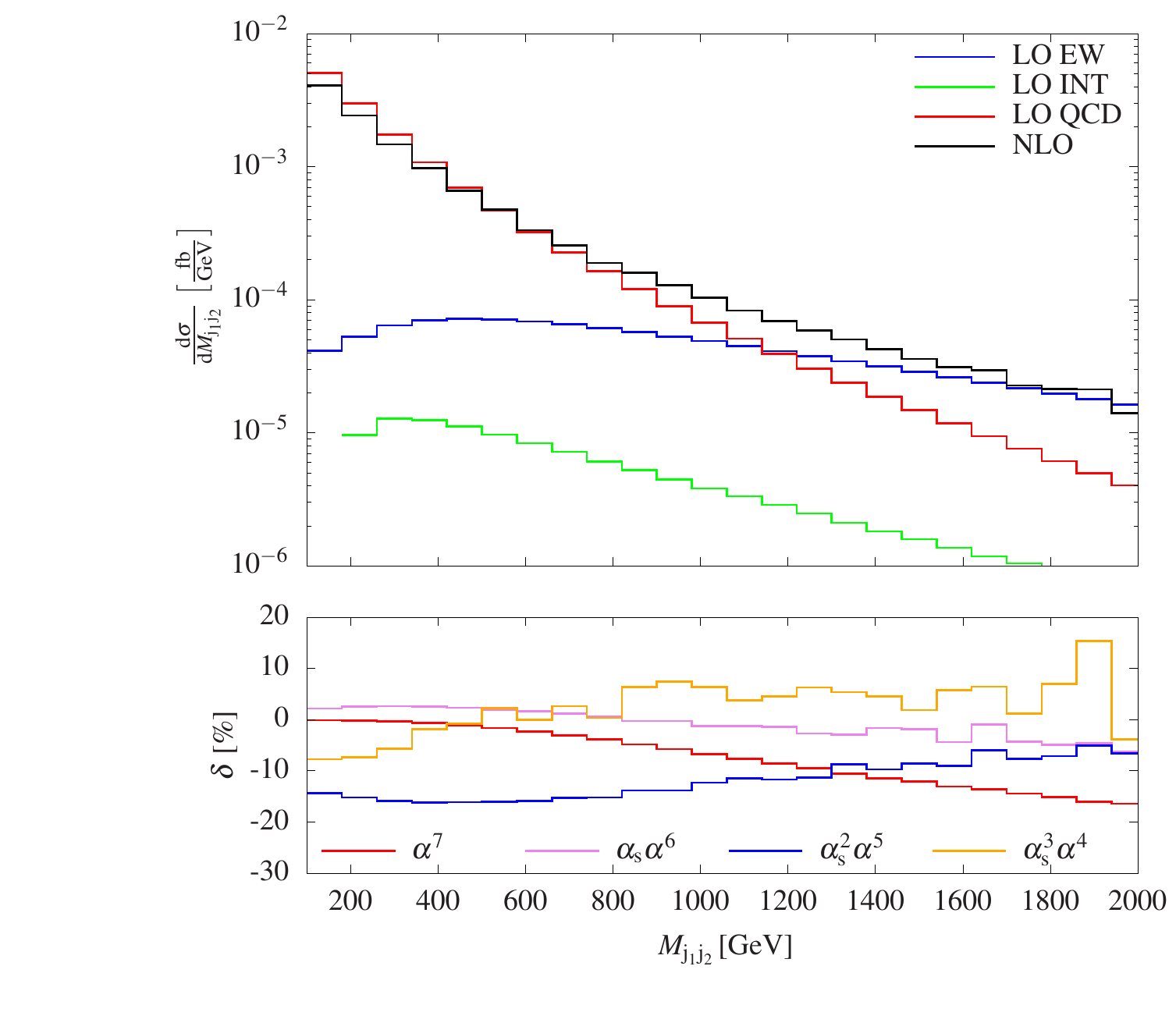}
\label{fig:mjj} 
\end{subfigure}
\begin{subfigure}{0.49\textwidth}
\centering
\subcaption{}
\includegraphics[width=1.\linewidth]{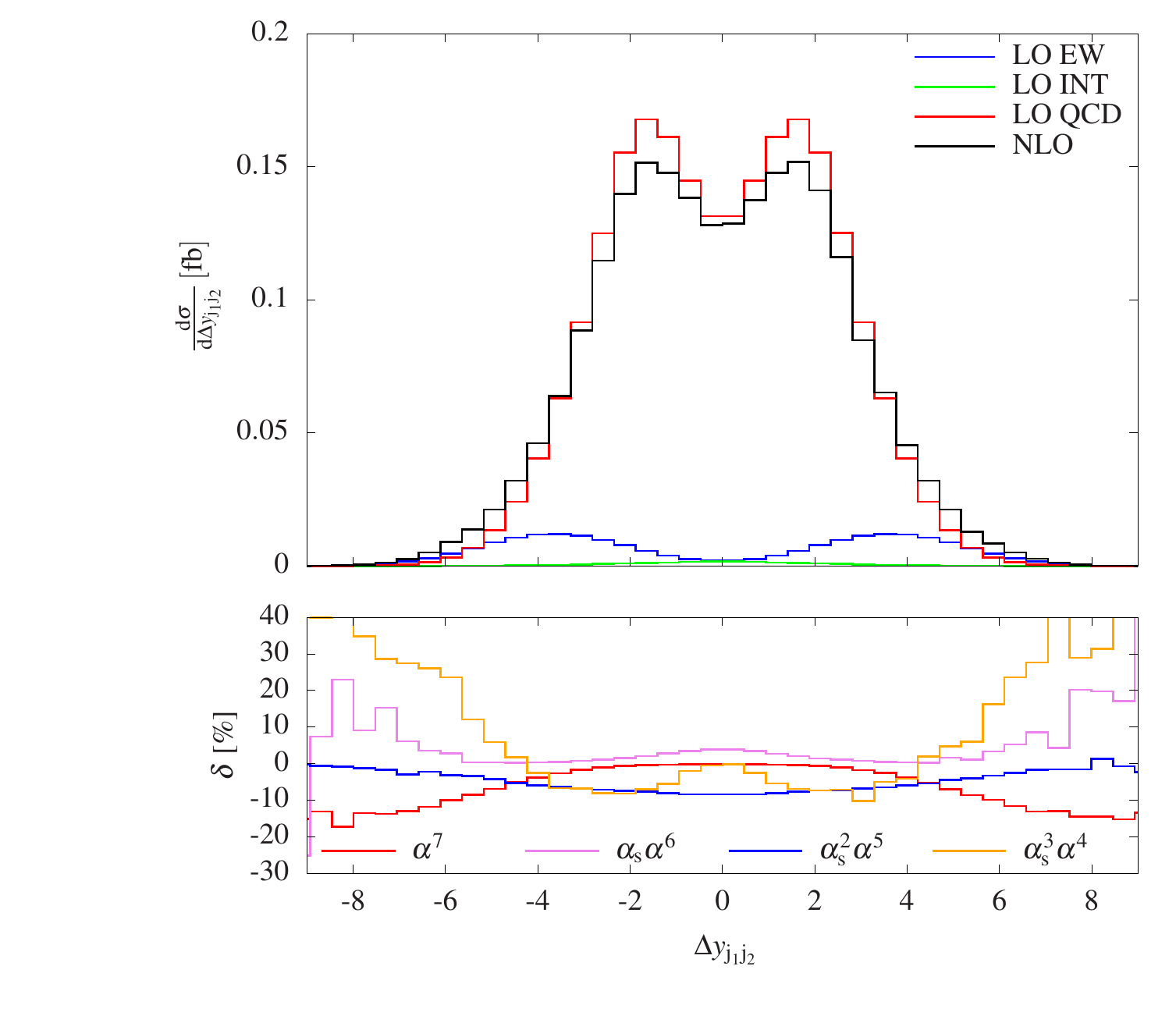}
\label{fig:dyjj}
\end{subfigure}%
\par\bigskip
\begin{subfigure}{0.49\textwidth}
\centering
\subcaption{}
\includegraphics[width=1.\linewidth]{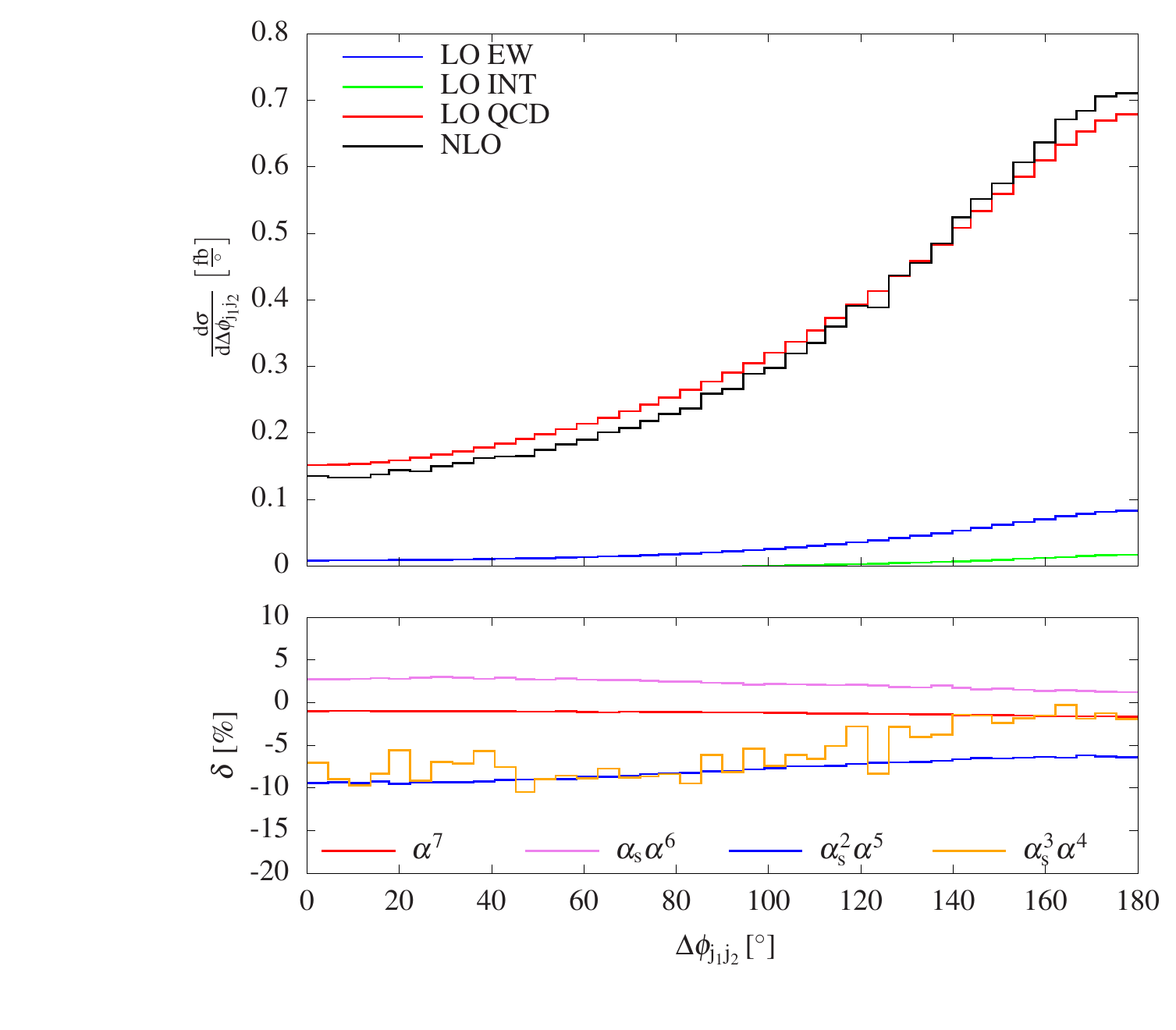}
\label{fig:phijj} 
\end{subfigure}
\begin{subfigure}{0.49\textwidth}
\centering
\subcaption{}
\includegraphics[width=1.\linewidth]{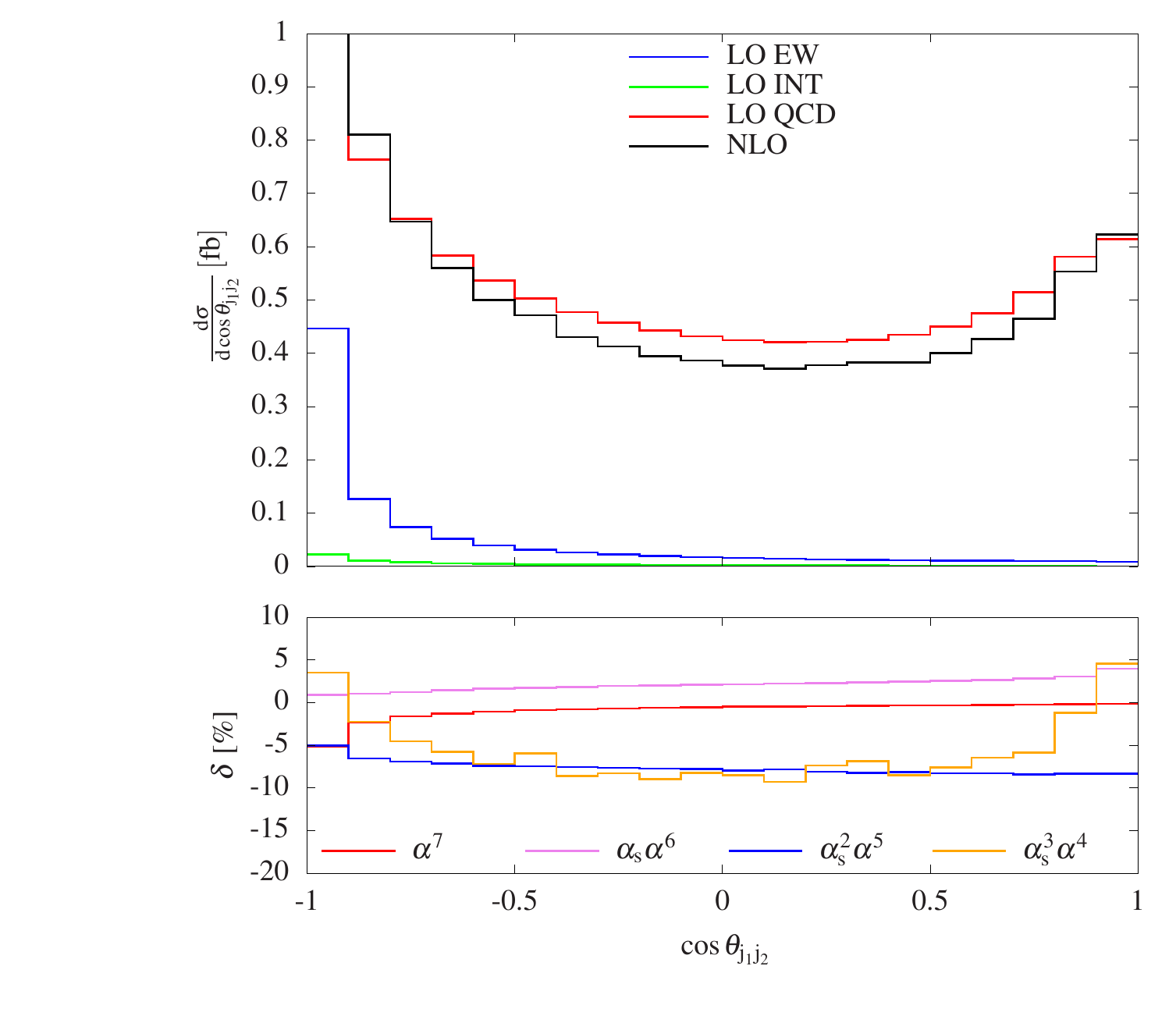}
\label{fig:cosjj}
\end{subfigure}%
\vspace*{-3ex}
\caption{Separate contributions of LO and NLO. The  upper panels show absolute predictions of orders
$\order{\alpha^6}$ (LO EW), $\order{\alphas\alpha^5}$ (LO INT),
$\order{\alphas^2\alpha^4}$ (LO QCD) and the complete NLO prediction. 
The lower panels display the contributions of orders 
$\order{\alpha^7}$, $\order{\alphas\alpha^6}$, $\order{\alphas^2\alpha^5}$, and
      $\order{\alphas^3\alpha^4}$  relative to the complete LO predictions.
The observables read as follows:
invariant mass of the two tagging jets (top left),
rapidity separation of the two tagging jets (top right),
azimuthal angle between the two tagging jets (bottom left), and
cosine of the angle between the two tagging jets (bottom right).}
\label{fig:NLOjj}
\end{figure}

\begin{figure}
\setlength{\parskip}{-4ex}
\begin{subfigure}{0.49\textwidth}
\centering
\subcaption{}
\includegraphics[width=1.\linewidth]{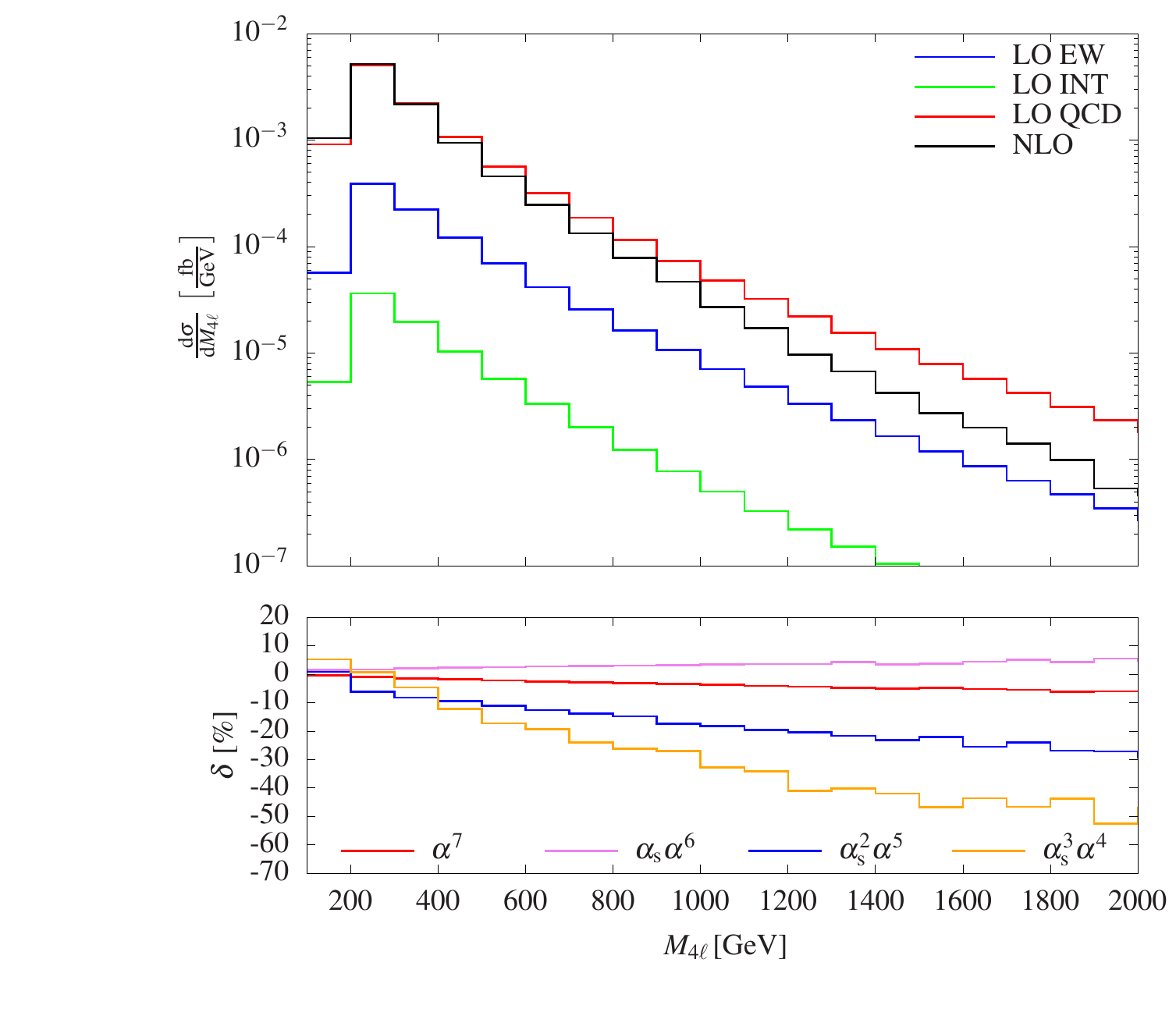}
\label{fig:m4l} 
\end{subfigure}
\begin{subfigure}{0.49\textwidth}
\centering
\subcaption{}
\includegraphics[width=1.\linewidth]{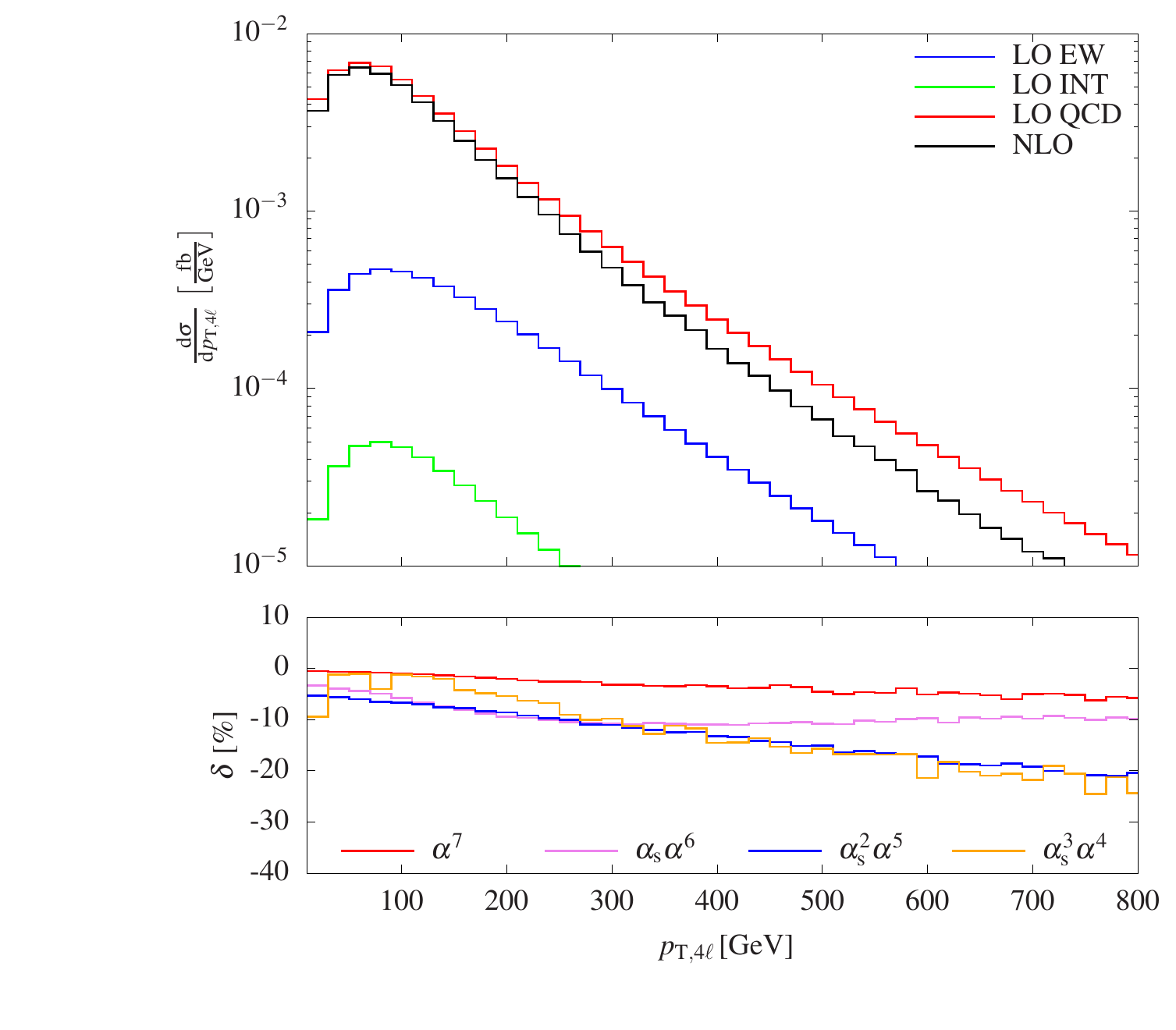}
\label{fig:pTzz} 
\end{subfigure}
\par\bigskip
\begin{subfigure}{0.49\textwidth}
\centering
\subcaption{}
\includegraphics[width=1.\linewidth]{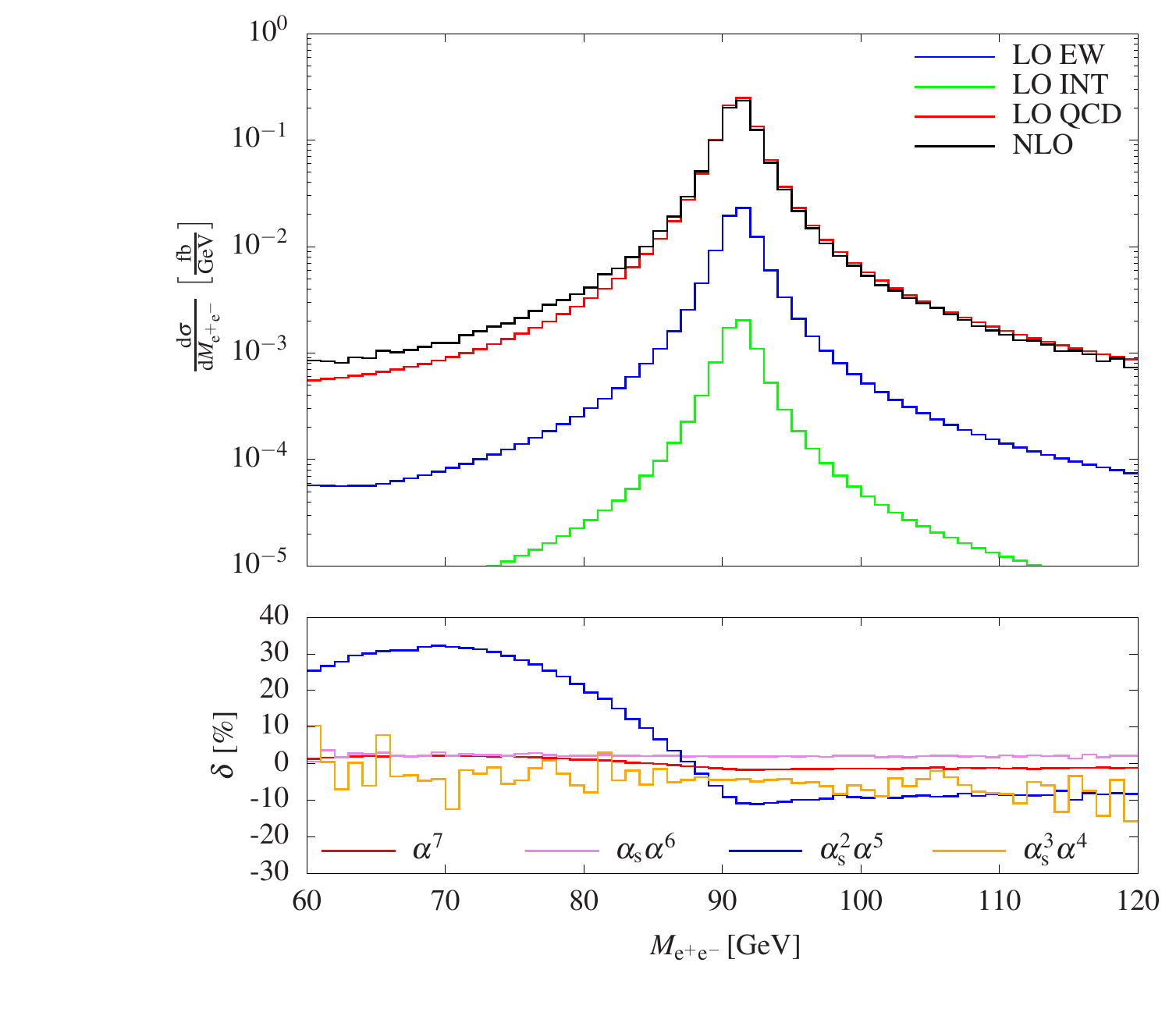}
\label{fig:mee}
\end{subfigure}%
\begin{subfigure}{0.49\textwidth}
\centering
\subcaption{}
\includegraphics[width=1.\linewidth]{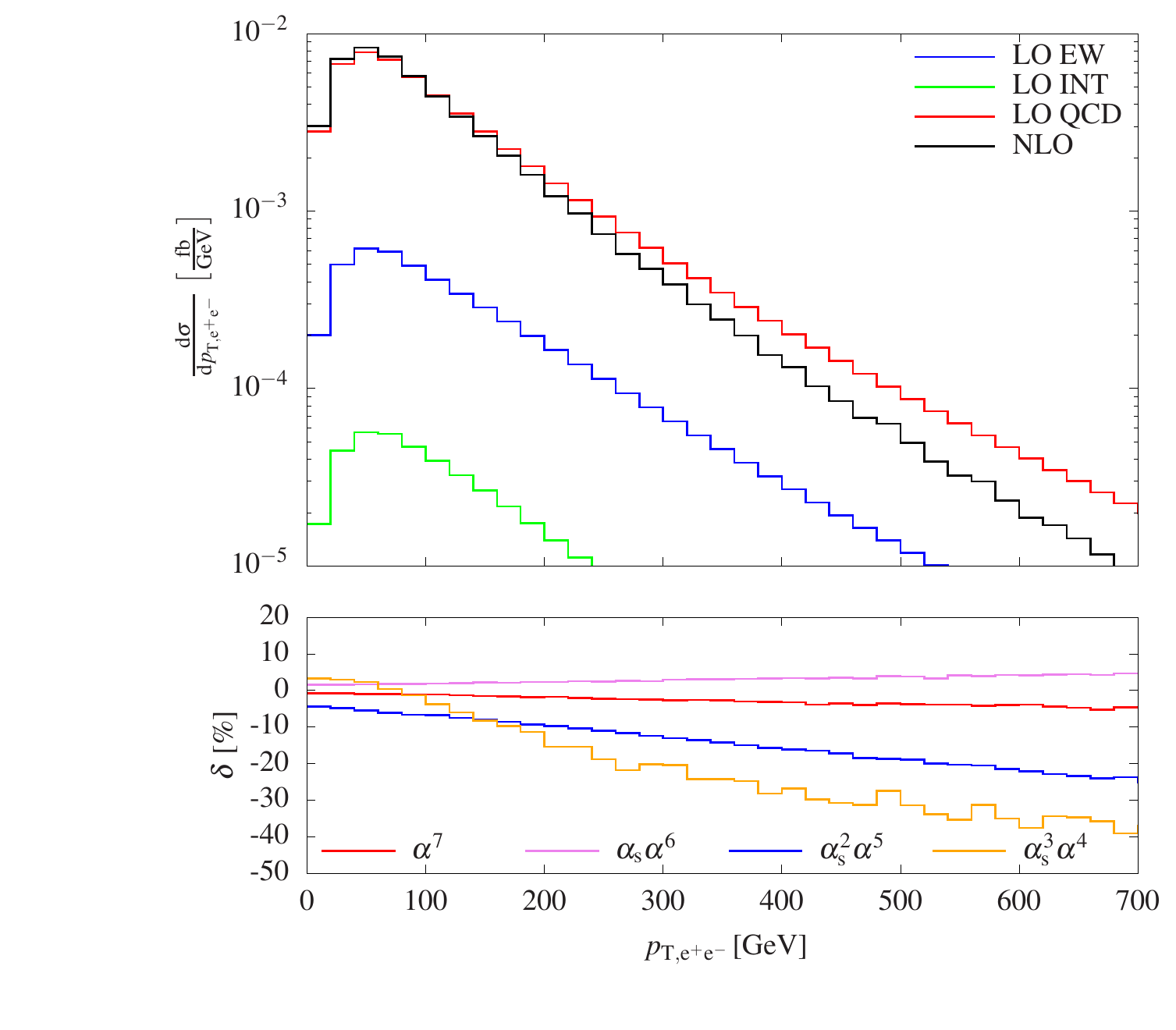}
\label{fig:pTee}
\end{subfigure}%
\vspace*{-3ex}
\caption{Same as for \reffi{fig:NLOjj} but for the observables:
invariant mass of four-lepton system (top left),
transverse momentum of the four leptons (top right),
invariant mass of the electron--positron system (bottom left), 
and
transverse momentum of the electron--positron system (bottom right).}
\label{fig:NLOZ}
\end{figure}

\begin{figure}
\setlength{\parskip}{-4ex}
\begin{subfigure}{0.49\textwidth}
\centering
\subcaption{}
\includegraphics[width=1.\linewidth]{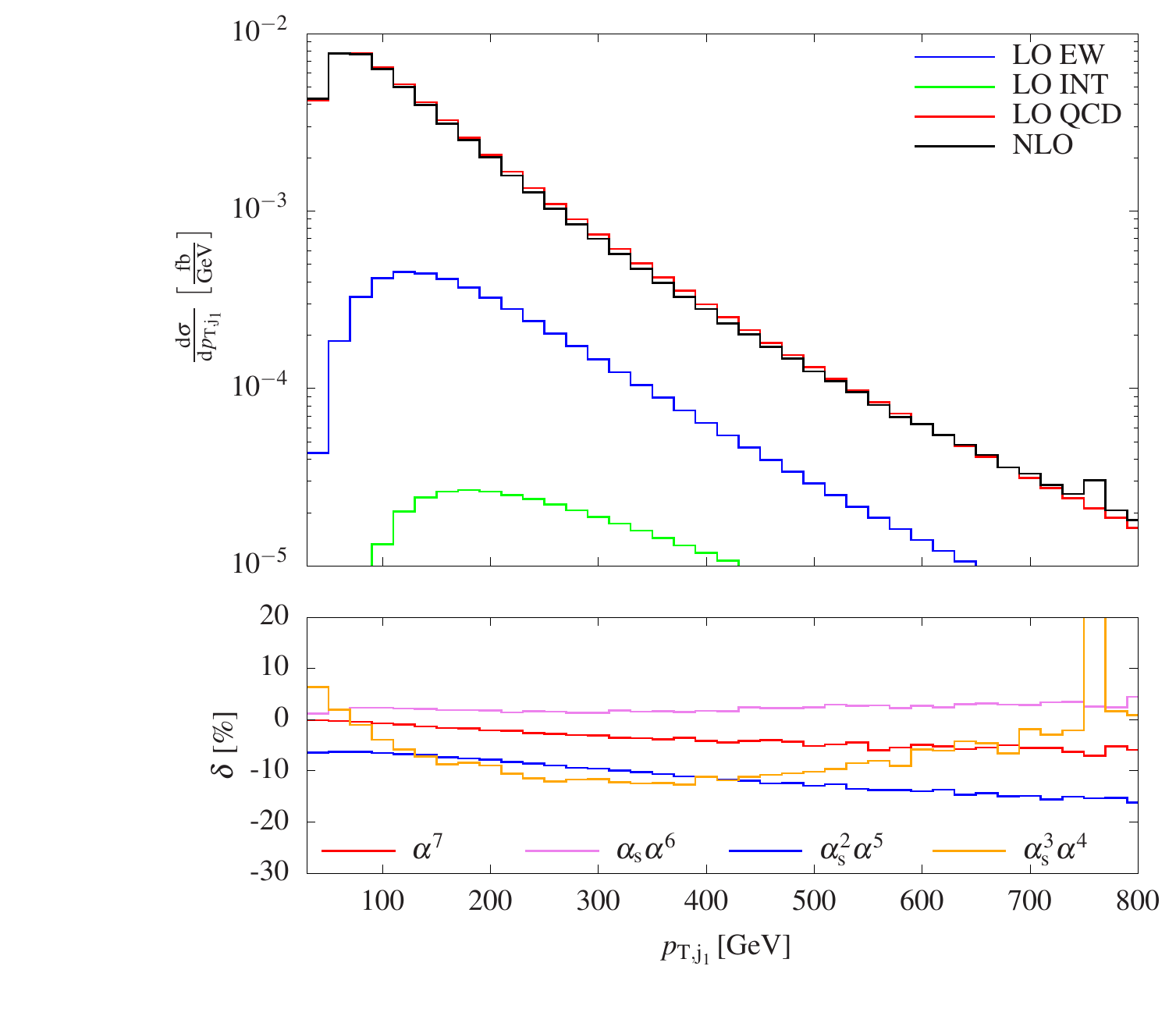}
\label{fig:pTj1} 
\end{subfigure}
\begin{subfigure}{0.49\textwidth}
\centering
\subcaption{}
\includegraphics[width=1.\linewidth]{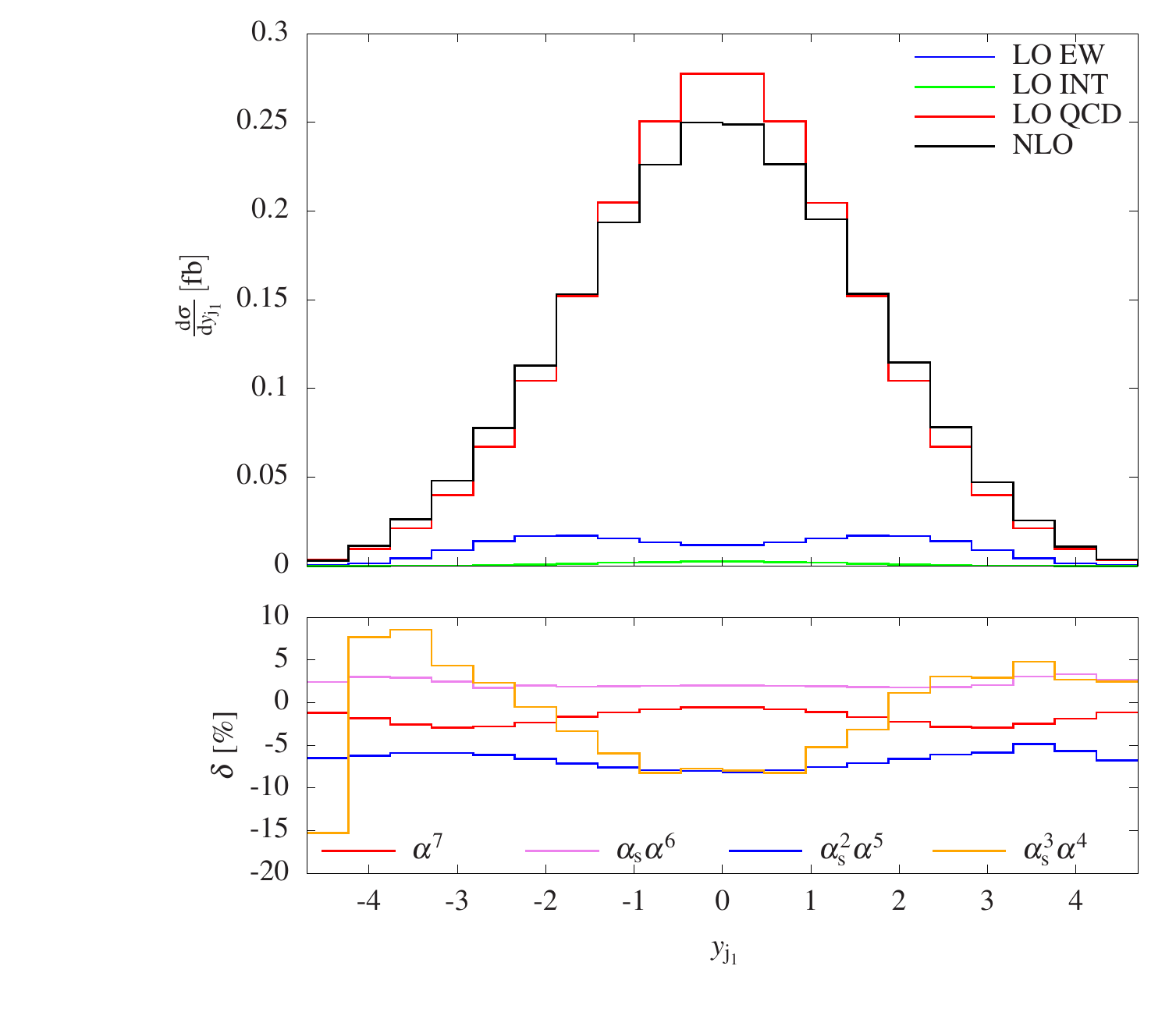}
\label{fig:yj1}
\end{subfigure}%
\par\bigskip
\begin{subfigure}{0.49\textwidth}
\centering
\subcaption{}
\includegraphics[width=1.\linewidth]{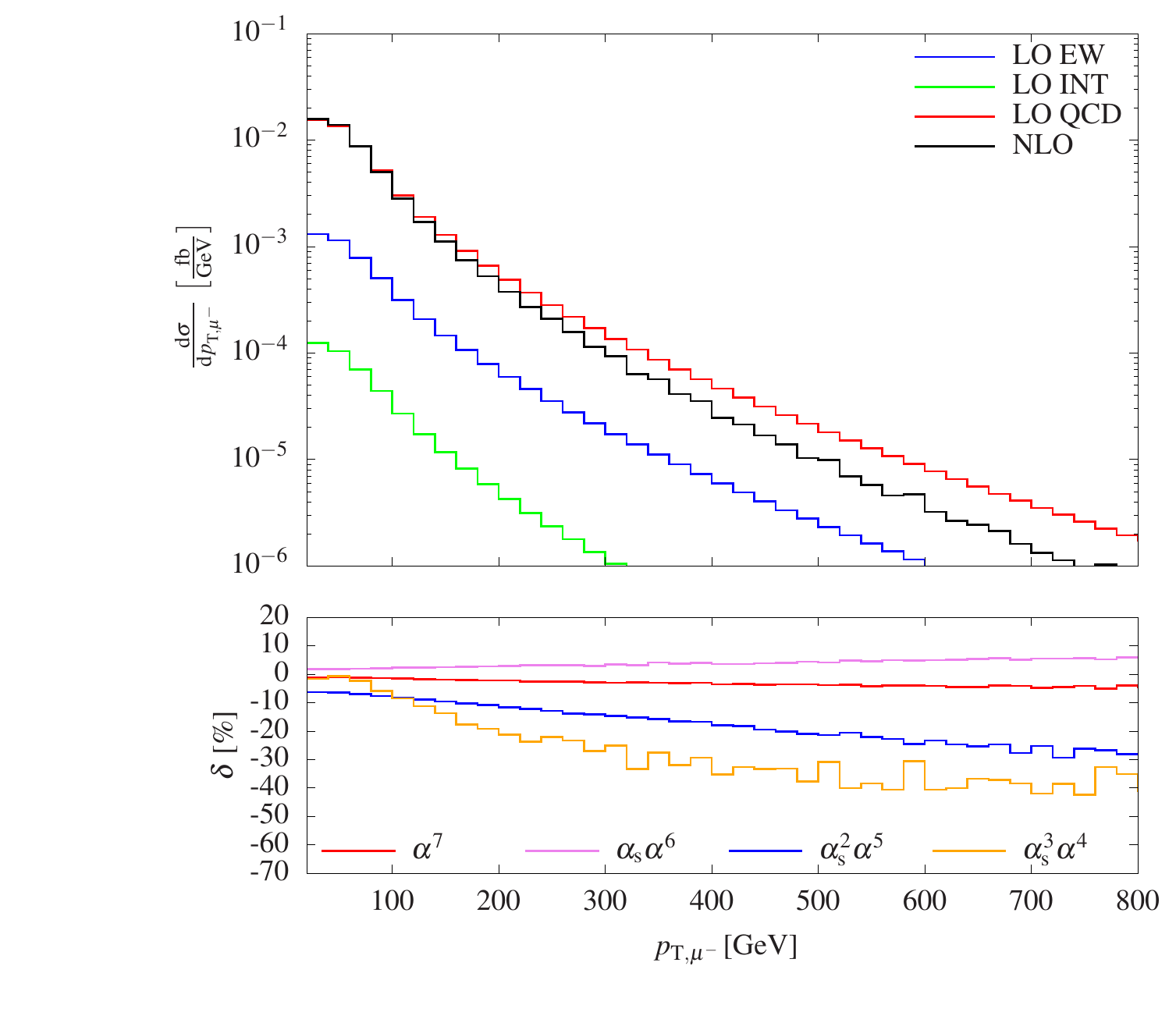}
\label{fig:pTmum} 
\end{subfigure}
\begin{subfigure}{0.49\textwidth}
\centering
\subcaption{}
\includegraphics[width=1.\linewidth]{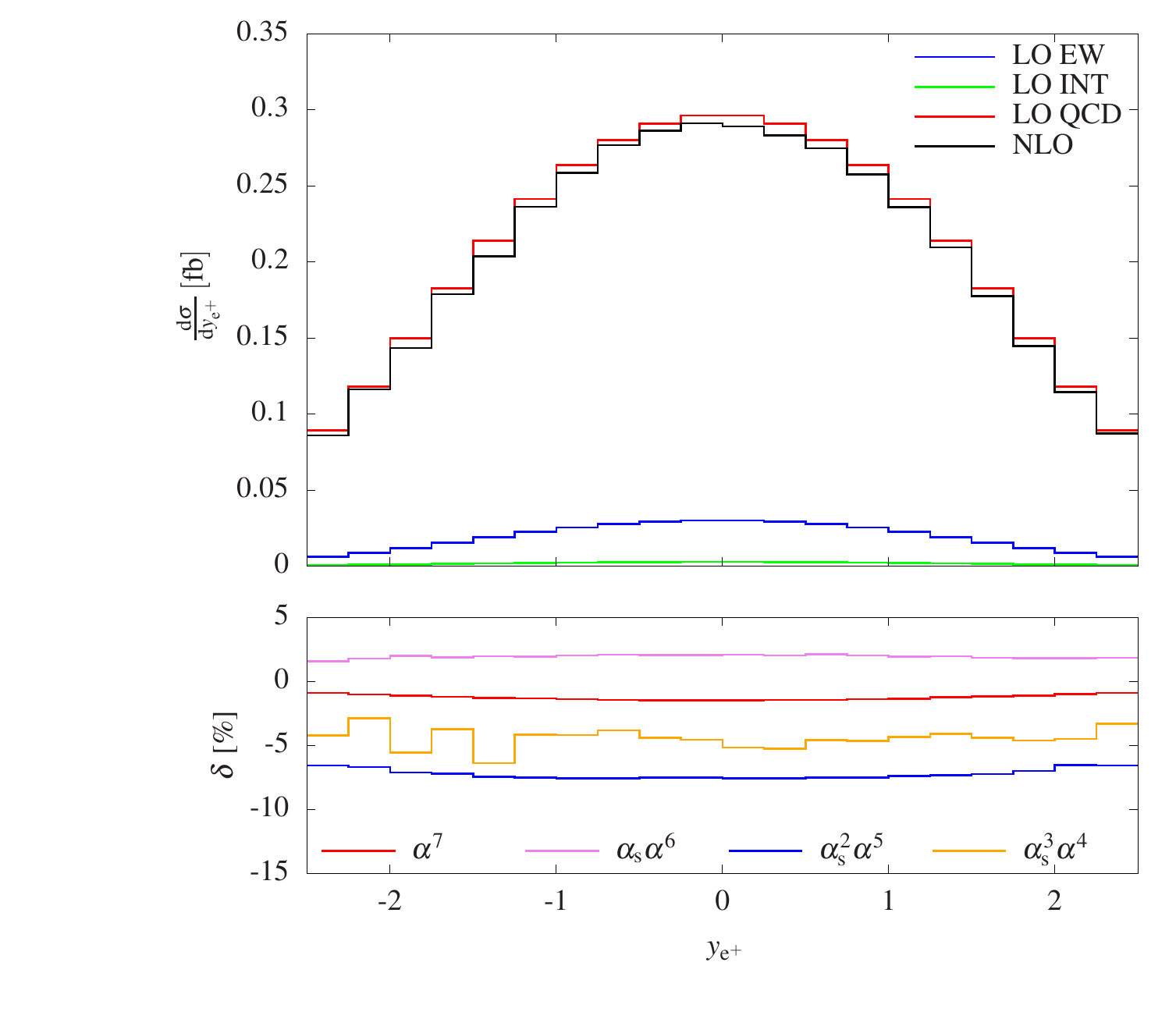}
\label{fig:yep}
\end{subfigure}%
\vspace*{-3ex}
\caption{Same as for \reffi{fig:NLOjj} but for the observables:
transverse momentum of the hardest jet (top left),
rapidity of the hardest jet (top right),
transverse momentum of the muon (bottom left),
and
rapidity of the positron (bottom right).
}
\label{fig:NLOje}
\end{figure}

\begin{figure}
\setlength{\parskip}{-4ex}
\begin{subfigure}{0.49\textwidth}
\centering
\subcaption{}
\includegraphics[width=1.\linewidth]{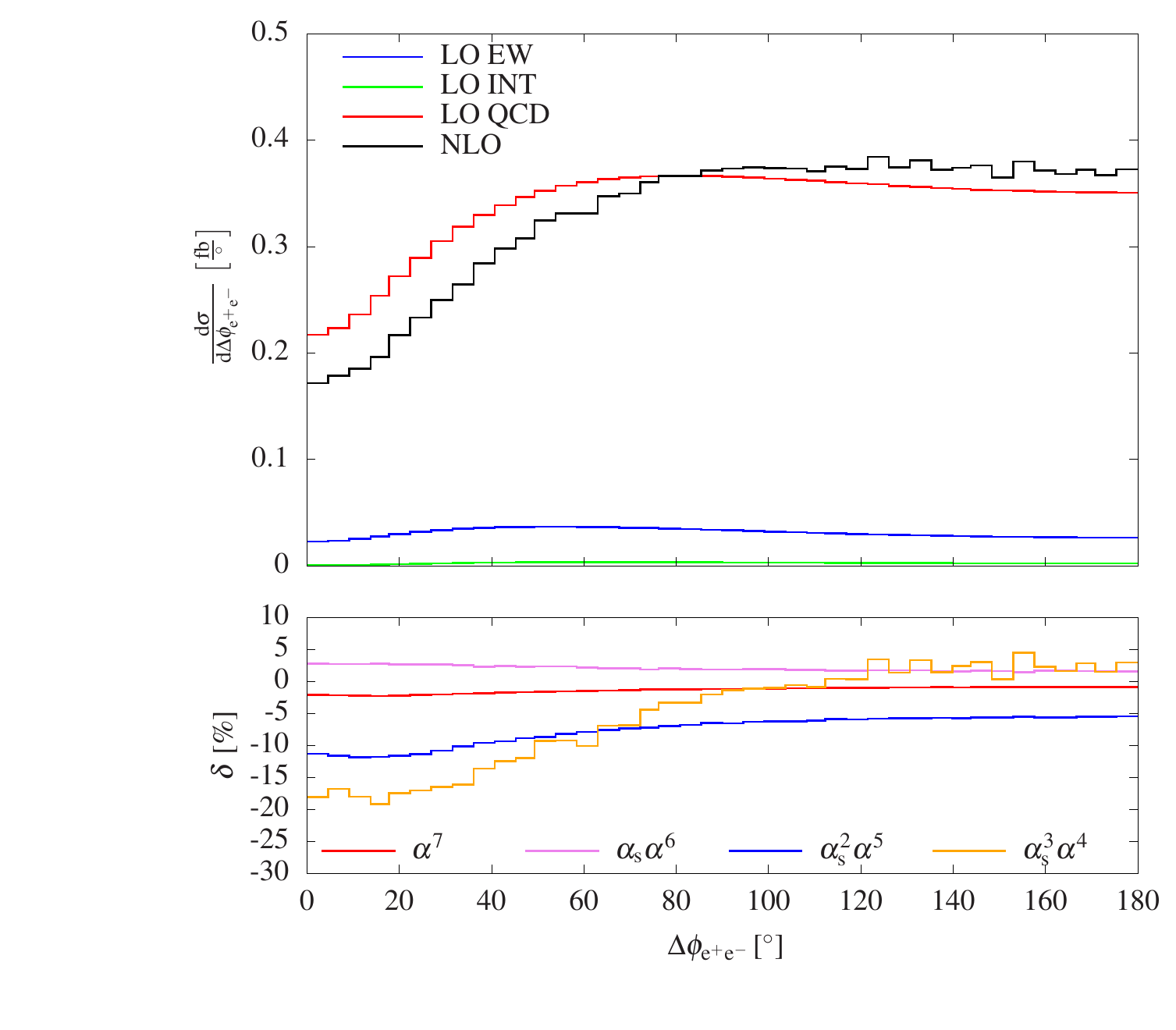}
\label{fig:aziee} 
\end{subfigure}
\begin{subfigure}{0.49\textwidth}
\centering
\subcaption{}
\includegraphics[width=1.\linewidth]{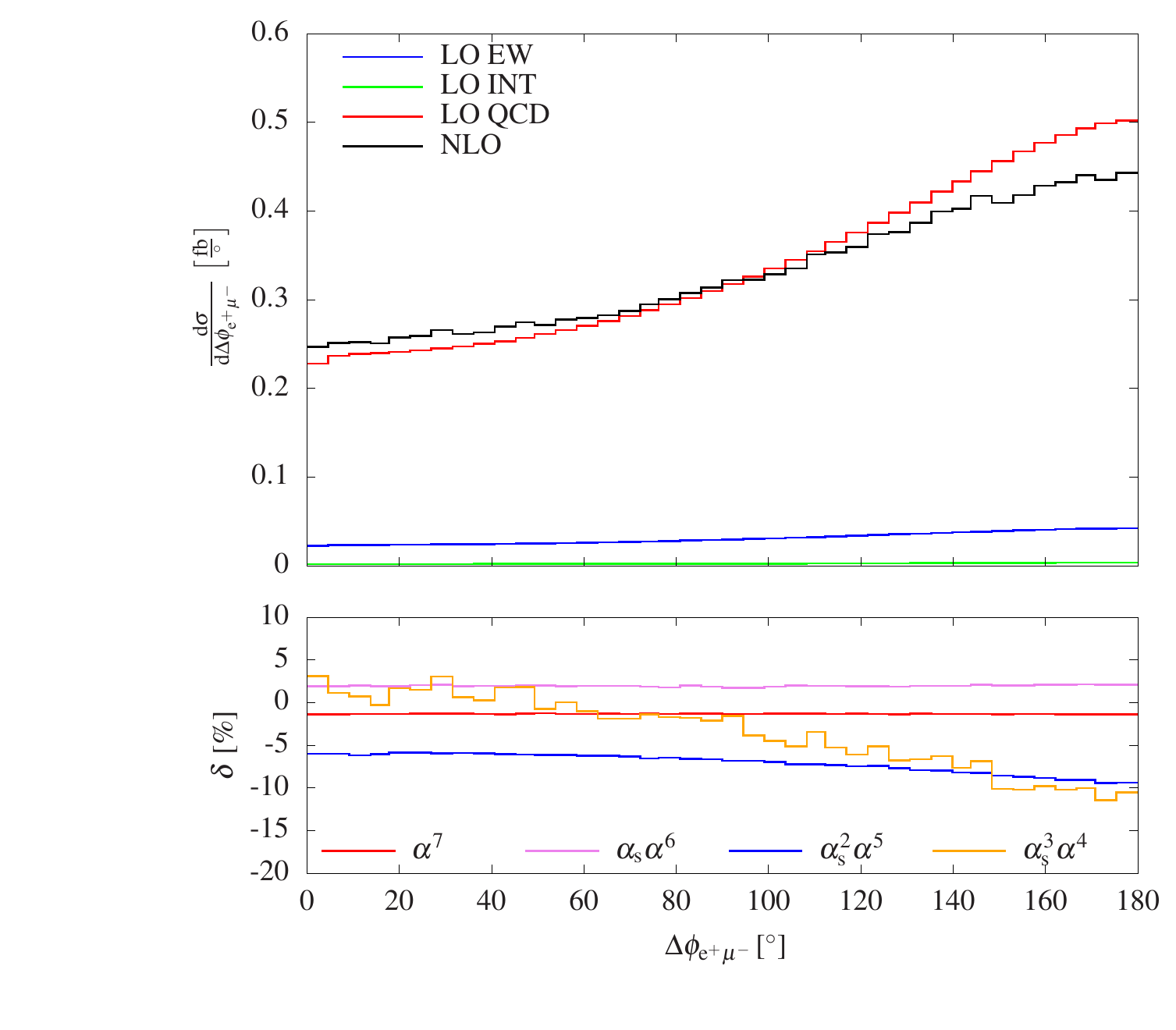}
\label{fig:aziem}
\end{subfigure}%
\par\bigskip
\begin{subfigure}{0.49\textwidth}
\centering
\subcaption{}
\includegraphics[width=1.\linewidth]{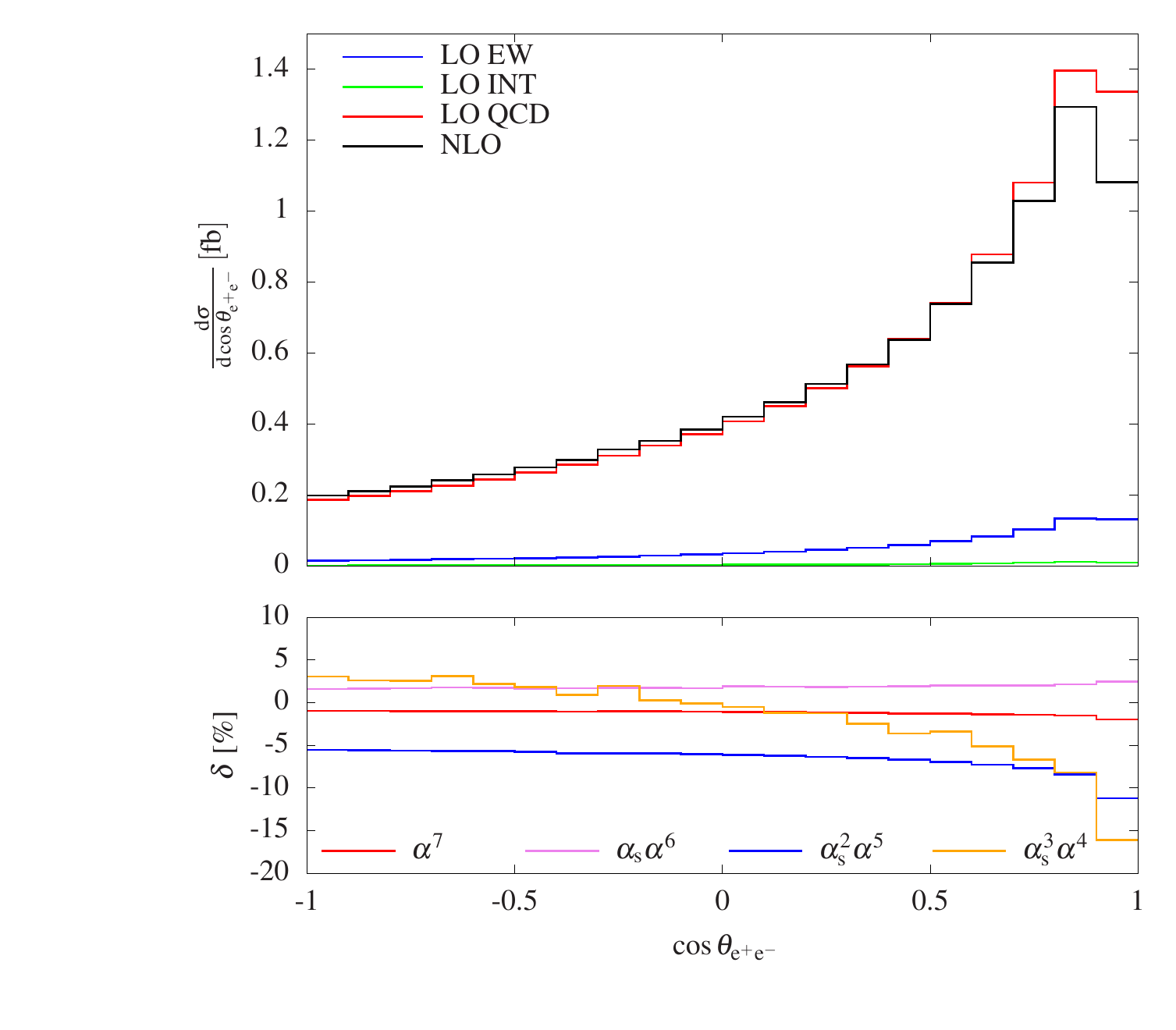}
\label{fig:cosee} 
\end{subfigure}
\begin{subfigure}{0.49\textwidth}
\centering
\subcaption{}
\includegraphics[width=1.\linewidth]{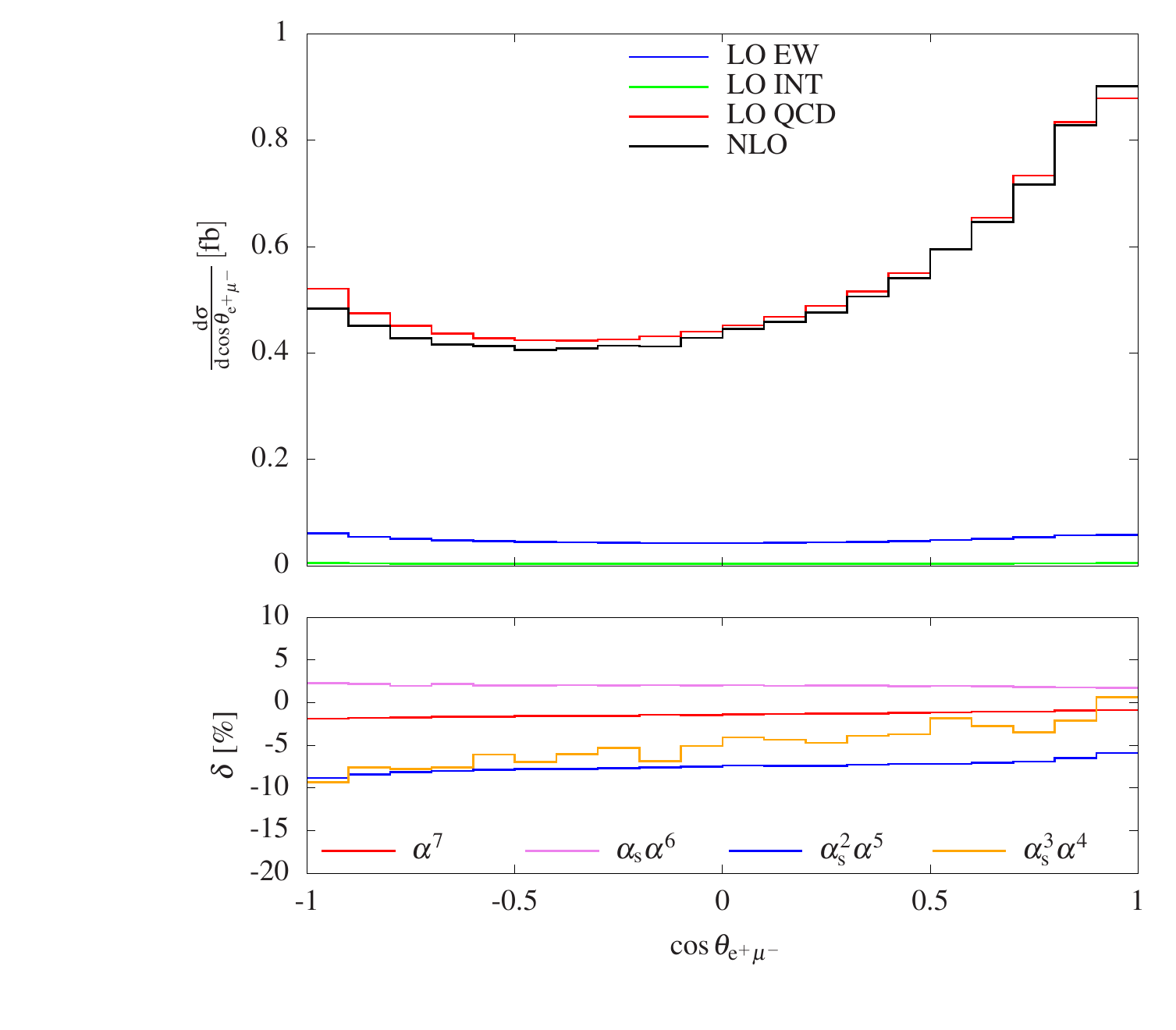}
\label{fig:cosem}
\end{subfigure}%
\vspace*{-3ex}
\caption{Same as for \reffi{fig:NLOjj} but for the observables:
azimuthal angle between the positron and the electron (top left),
azimuthal angle between the positron and the muon (top right),
cosine of the angle between the positron and the electron (bottom left), and
cosine of the angle between the positron and the muon (bottom right).}
\label{fig:NLOll}
\end{figure}

In this section, results for differential distributions at $13\TeV$ are discussed
for the inclusive setup defined in \refse{se:eventselection}.  In the
first part, we display the contributions of the different orders
separately. In the upper panels of
\reffis{fig:NLOjj}--\ref{fig:NLOll}, all LO contributions of orders
$\order{\alpha^6}$, $\order{\alphas\alpha^5}$,  and $\order{\alphas^2\alpha^4}$ are shown together
with the complete NLO predictions. In the lower panels, the NLO
corrections of orders $\order{\alpha^7}$, $\order{\alphas\alpha^6}$,
$\order{\alphas^2\alpha^5}$, and $\order{\alphas^3\alpha^4}$ are
plotted separately, normalised to
the complete LO predictions.  In the second part of this section we
present full LO and NLO predictions with the corresponding scale
uncertainty.

The first set of differential observables in \reffi{fig:NLOjj} relates to the two tagging jets as defined in Eqs.~\eqref{eq:jet1} and \eqref{eq:vbscuts}.
The distributions in the invariant mass (\reffi{fig:mjj}) and the
rapidity separation (\reffi{fig:dyjj}) of the two tagging jets are
typically used to improve the ratio of the EW component over the QCD
one.  These two distributions differ substantially from all others 
shown in this paper, since they receive sizeable contributions
from the order $\order{\alpha^6}$ for large $M_{\Pj_1\Pj_2}$ or large $\Delta
y _{\Pj_1\Pj_2}$, as can be seen in the upper part of both plots,
while all other distributions are dominated by the order
$\order{\alphas^2\alpha^4}$ throughout. Thus, the normalisation of the
relative corrections is dominated by the $\order{\alphas^2\alpha^4}$
contributions for small $M_{\Pj_1\Pj_2}$ and $\Delta
y _{\Pj_1\Pj_2}$, but by the $\order{\alpha^6}$ ones for large variables.
Owing to this varying normalisation,
the EW corrections of order $\order{\alpha^7}$ are large for large
$M_{\Pj_1\Pj_2}$ or large $\Delta y _{\Pj_1\Pj_2}$ (reaching $-18\%$
at $M_{\Pj_1\Pj_2}=2\TeV$) and small otherwise.
The normalisation also explains the opposite behaviour of the (EW) corrections of order $\order{\alphas^2\alpha^5}$,
which reach  about $-15\%$ at $M_{\Pj_1\Pj_2}=400\GeV$ but are reduced to
about $-5\%$ to $-7\%$ at $2\TeV$ in the invariant-mass distribution. 
Despite the fact that these large EW corrections can be traced back to
Sudakov logarithms, they become relatively smaller at high energies as
the LO contribution of order $\order{\alphas^2\alpha^4}$ (to which
these corrections act on) is suppressed there. 
The corrections of QCD type, $\order{\alphas\alpha^6}$ and
$\order{\alphas^3\alpha^4}$, stay within $\pm10\%$ apart from the
region of large rapidity separations.  In particular, the QCD
corrections of order $\mathcal{O}(\alphas^3\alpha^4)$ turn very large
(over $+40\%$) there, but this part of the phase space is rather
suppressed. These QCD corrections are positive for invariant masses
above $M_{\Pj_1\Pj_2}=500\GeV$ and tend to increase towards higher
invariant masses.  

The corrections to the distributions in the azimuthal angle (\reffi{fig:phijj}) and the
cosine of the angle between the two tagging jets (\reffi{fig:cosjj}) show
rather mild variations that do not exceed $10\%$
over the full phase-space range. The EW corrections of order
$\mathcal{O}(\alpha^7)$ vary by less than $2\%$ with the exception of
$\cos\theta_{\Pj_1\Pj_2}\approx-1$.  The EW corrections of order
$\mathcal{O}(\alphas^2\alpha^5)$ are almost $-10\%$ for small angles
between the jets, while they decrease in magnitude to $-5\%$ for large
angles.  The pure QCD corrections of order
$\mathcal{O}(\alphas^3\alpha^4)$ grow in the azimuthal-angle
distribution from $-10\%$ up to almost $0\%$, while they are minimal
for small values of $\cos\theta_{\Pj_1\Pj_2}$ and maximal for $\cos\theta_{\Pj_1\Pj_2}=\pm1$.

In \reffi{fig:NLOZ}, several leptonic observables are shown,
including distributions in the invariant mass (\reffi{fig:m4l})
and transverse momentum (\reffi{fig:pTzz}) of the four leptons as well
as in the invariant mass (\reffi{fig:mee}) and the transverse
momentum (\reffi{fig:pTee}) of the electron--positron pair.
The distributions in the invariant mass and transverse momentum of the
four leptons and the transverse momentum of the electron--positron
pair display qualitatively rather similar features as they are
correlated.  In absolute, they all reach a maximum at low values to
decrease steeply towards high energy.  The
$\mathcal{O}(\alphas^2\alpha^5)$ corrections show the typical negative
increase owing to Sudakov logarithms reaching up to $-30\%$ in the
considered kinematical range. The corresponding behaviour of the
$\mathcal{O}(\alpha^7)$ corrections is damped by the normalisation to
LO including the dominating QCD contributions.  The QCD corrections of
order $\mathcal{O}(\alphas^3\alpha^4)$ show an even stronger impact
towards high energy scales reaching $-50\%$ at $M_{4\ell}=2\TeV$ and
$-40\%$ at $p_{\rT,\Pe^+\Pe^-}=700\GeV$. This is caused by our choice of
the renormalisation scale \refeq{eq:defscale} that is adapted to the
VBS contributions of order $\mathcal{O}(\alpha^6)$. For large energy
scales in leptonic variables this scale is too small leading to a too
large LO cross section and thus large negative QCD corrections. Note
that the $\mathcal{O}(\alphas\alpha^6)$ corrections, \ie the QCD
corrections to the VBS contributions, stay within $\pm10\%$ for all 
distributions in \reffi{fig:NLOZ}. The distribution in the invariant
mass of the electron--positron pair (\reffi{fig:mee}) is characterised
by the Z-boson resonance at $\sim91\GeV$ in all LO contributions.
While the QCD corrections are rather flat across the resonance, the EW
corrections of order $\mathcal{O}(\alphas^2\alpha^5)$ exhibit a
pronounced radiative tail below the resonance. The corrections due to
final-state photon radiation exceed $+30\%$ around $70\GeV$. The
corresponding radiative tail is also present in the corrections of
order  $\mathcal{O}(\alpha^7)$ but suppressed by one order of
magnitude by the LO QCD contributions.

In  \reffi{fig:NLOje} we present distributions related to the
leading jet or a single lepton.
In the distribution in the transverse momentum of the hardest jet (\reffi{fig:pTj1}), the
corrections of order $\mathcal{O}(\alphas^3\alpha^4)$ show the most
interesting behaviour.  They start around $6\%$ at $30\GeV$, reach a
minimum of about $-12\%$ around $350\GeV$ to finally grow to $1\%$ at
$800\GeV$. The moderate rise of these corrections towards high
transverse momenta can be attributed to the choice \ref{eq:defscale}
of renormalisation scale.  The 
increase of the QCD corrections towards small $p_{\rT,\Pj_1}$ has
already been observed in other VBS/VBF processes
\cite{Biedermann:2017bss,Denner:2019tmn,Denner:2020zit,Dreyer:2020xaj}
and is due to the real QCD radiation.  The corrections of order
$\mathcal{O}(\alphas^2\alpha^5)$ decrease steadily towards high energy
by about $10\%$.  Regarding the rapidity distribution of the hardest
jet (\reffi{fig:yj1}), the corrections of order
$\mathcal{O}(\alphas^2\alpha^5)$ vary only by few per cent in the
whole kinematic range.  The $\mathcal{O}(\alphas^3\alpha^4)$
corrections, on the other hand, go from $+8\%$ in the peripheral
region down to $-7\%$ in the central one.  The differential
distributions in the transverse momentum of the muon
(\reffi{fig:pTmum}) and the rapidity of the positron (\reffi{fig:yep})
are rather standard.  The $p_{\rT,\mu^-}$ distribution behaves like
other transverse-momentum distributions of leptons shown above, while
the variation of all corrections to the $y_{\Pe^+}$ distribution is below
about two per cent.

Next we consider leptonic angular distributions in
\reffi{fig:NLOll}.
The fraction of the LO EW contribution is somewhat enhanced for small
and moderate $\cos\theta_{\Pe^+\mu^-}$.
In all distributions of \reffi{fig:NLOll}, the corrections of orders $\mathcal{O}(\alpha^7)$ and
$\mathcal{O}(\alphas\alpha^6)$ are small, inheriting their value from
the one of the fiducial cross section.  The distortion of the LO
distributions results mostly from the $\mathcal{O}(\alphas^3\alpha^4)$
corrections and to some extent from the
$\mathcal{O}(\alphas^2\alpha^5)$ corrections. Both types of corrections are
largest for small angles between the electron--positron pairs and
large angles between the muon--positron pairs, which typically appear
for events at large energies. Thus, the variation of these corrections
is driven by EW Sudakov logarithms and the choice of the QCD
renormalisation scale. The corrections of order
$\mathcal{O}(\alphas^3\alpha^4)$ vary between $-20\%$ and $+3\%$ for
different angles
between electron--positron pairs and between $-10\%$ and $+3\%$ with
angles between muon--positron pairs. The corresponding variations of the
$\mathcal{O}(\alphas^2\alpha^5)$ corrections are from $-12\%$ to
$-5\%$ and from $-9\%$ to $-6\%$, respectively.

\begin{figure}
\setlength{\parskip}{-4ex}
\begin{subfigure}{0.49\textwidth}
\centering
\subcaption{}
\includegraphics[width=1.\linewidth]{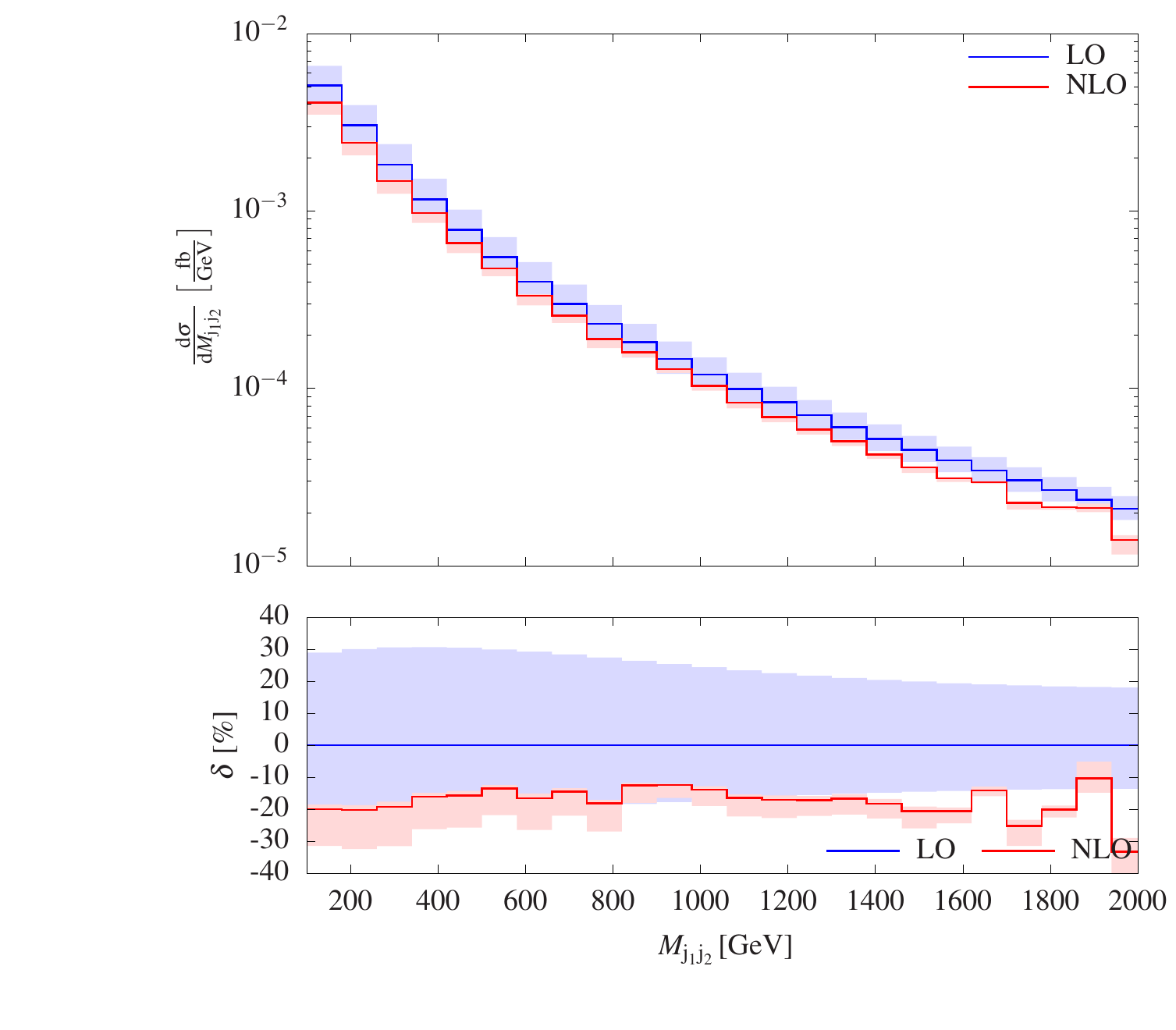}
\label{fig:mjjs} 
\end{subfigure}
\begin{subfigure}{0.49\textwidth}
\centering
\subcaption{}
\includegraphics[width=1.\linewidth]{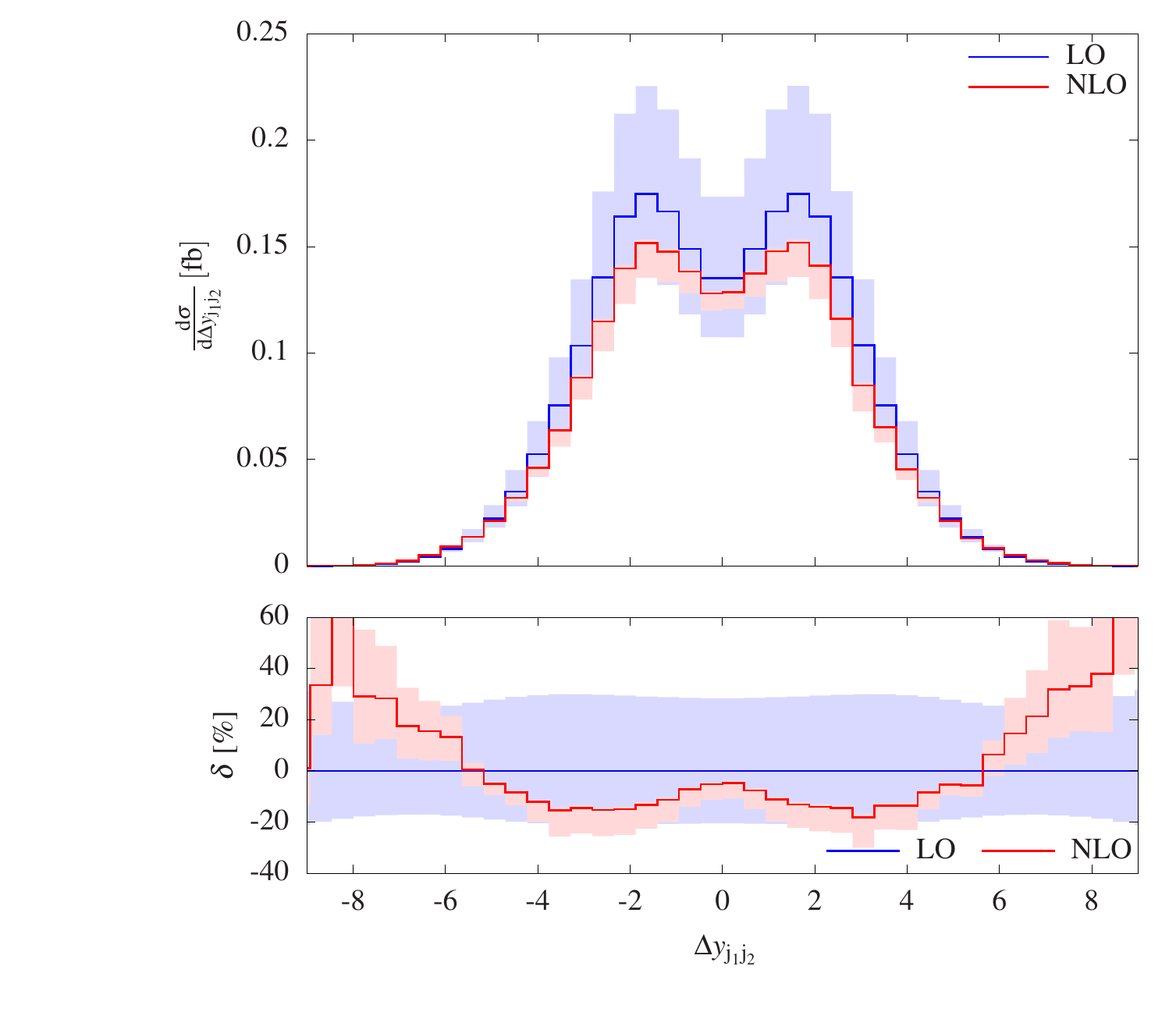}
\label{fig:dyjjs}
\end{subfigure}%
\par\bigskip
\begin{subfigure}{0.49\textwidth}
\centering
\subcaption{}
\includegraphics[width=1.\linewidth]{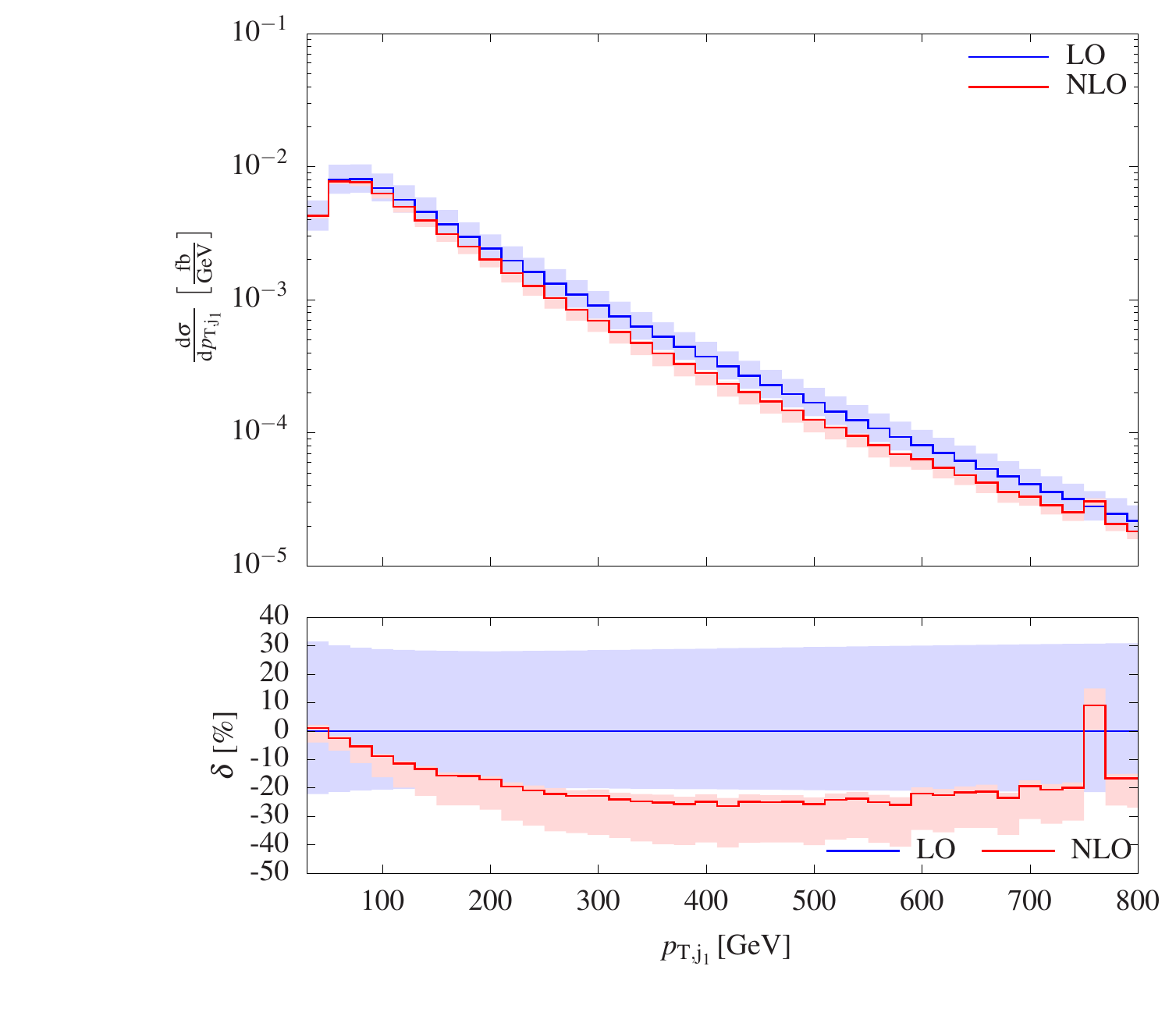}
\label{fig:ptj1s} 
\end{subfigure}
\begin{subfigure}{0.49\textwidth}
\centering
\subcaption{}
\includegraphics[width=1.\linewidth]{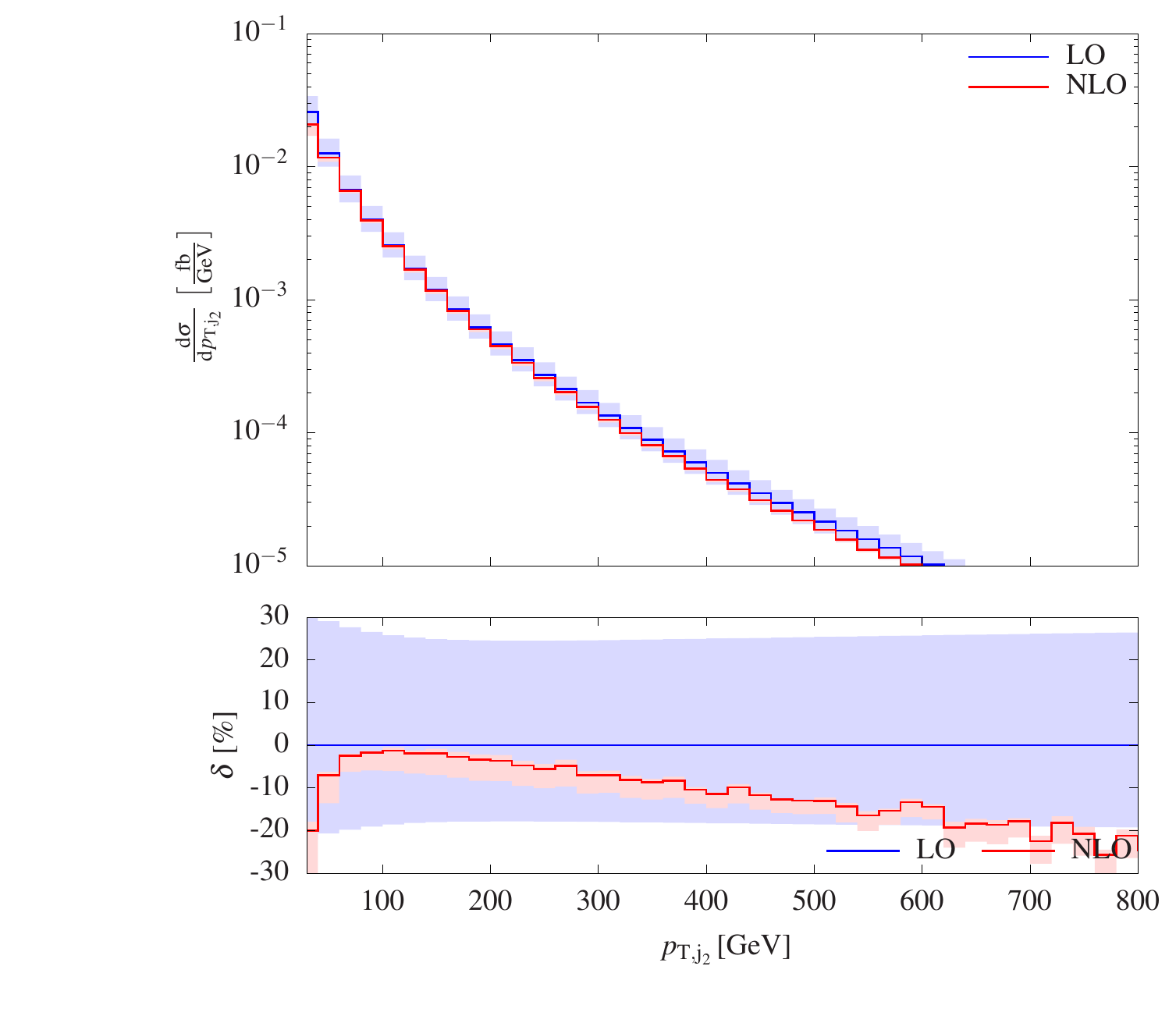}
\label{fig:ptj2s}
\end{subfigure}%
\vspace*{-3ex}
\caption{Differential distributions including 7-point
  scale uncertainties.
The upper panels show absolute predictions for LO and NLO while the lower ones show
relative NLO corrections with respect to the LO predictions at the
central scale and the relative LO scale uncertainty. 
The full LO predictions include orders $\mathcal{O}(\alpha^6)$, $\mathcal{O}(\alphas\alpha^5)$, and $\mathcal{O}(\alphas^2\alpha^4)$,
while the NLO ones comprise $\mathcal{O}(\alpha^7)$, $\mathcal{O}(\alphas\alpha^6)$, 
$\mathcal{O}(\alphas^2\alpha^5)$, and $\mathcal{O}(\alphas^3\alpha^4)$ contributions.
The observables read as follows:
invariant mass of the two tagging jets (top left),
rapidity difference between the two tagging jets (top right),
transverse momentum of the hardest jet (bottom left), and
transverse momentum of the second hardest jet (bottom right).}
\label{fig:NLO_scale}
\end{figure}

\begin{figure}
\setlength{\parskip}{-4ex}
\begin{subfigure}{0.49\textwidth}
\centering
\subcaption{}
\includegraphics[width=1.\linewidth]{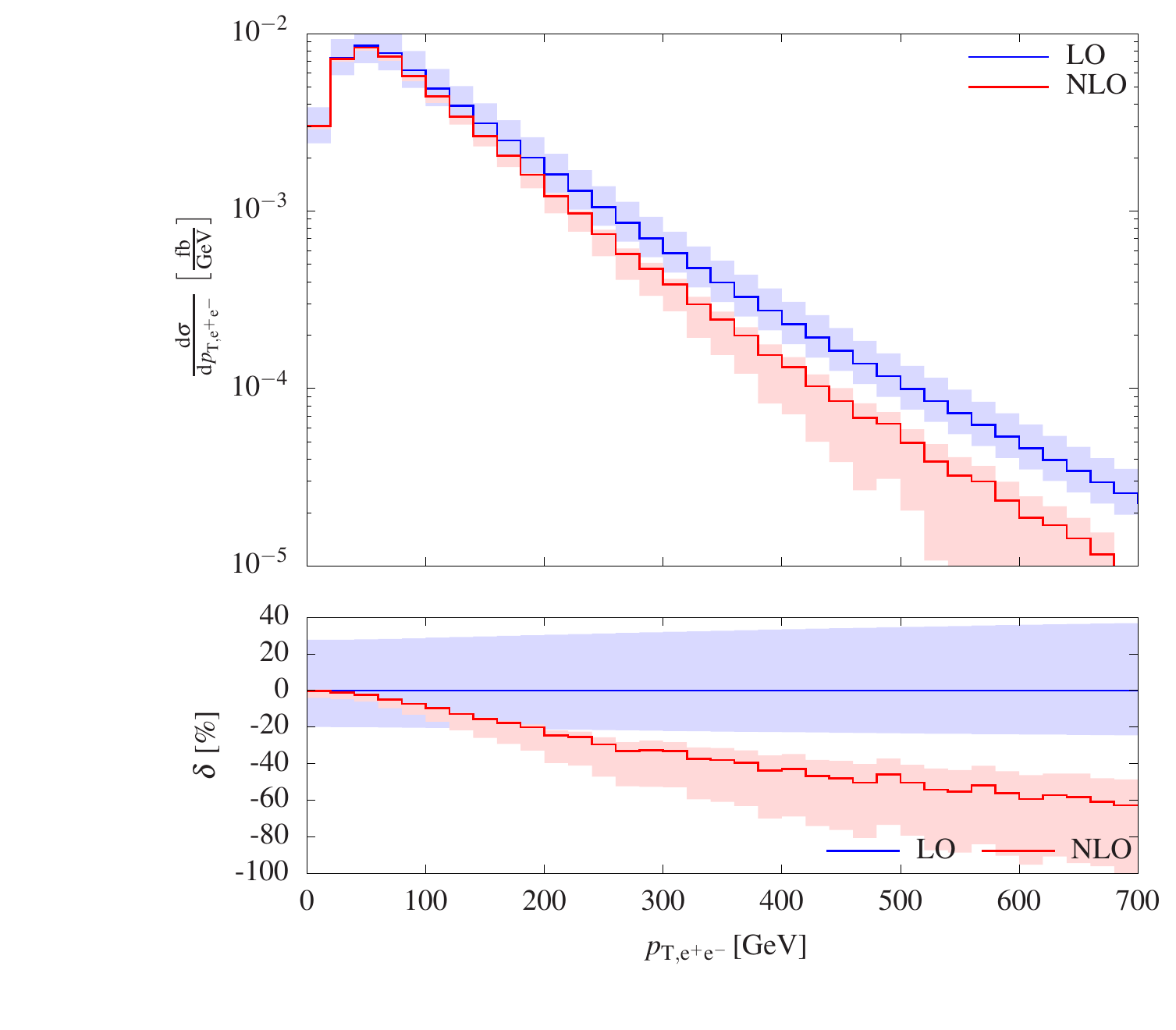}
\label{fig:pTees} 
\end{subfigure}
\begin{subfigure}{0.49\textwidth}
\centering
\subcaption{}
\includegraphics[width=1.\linewidth]{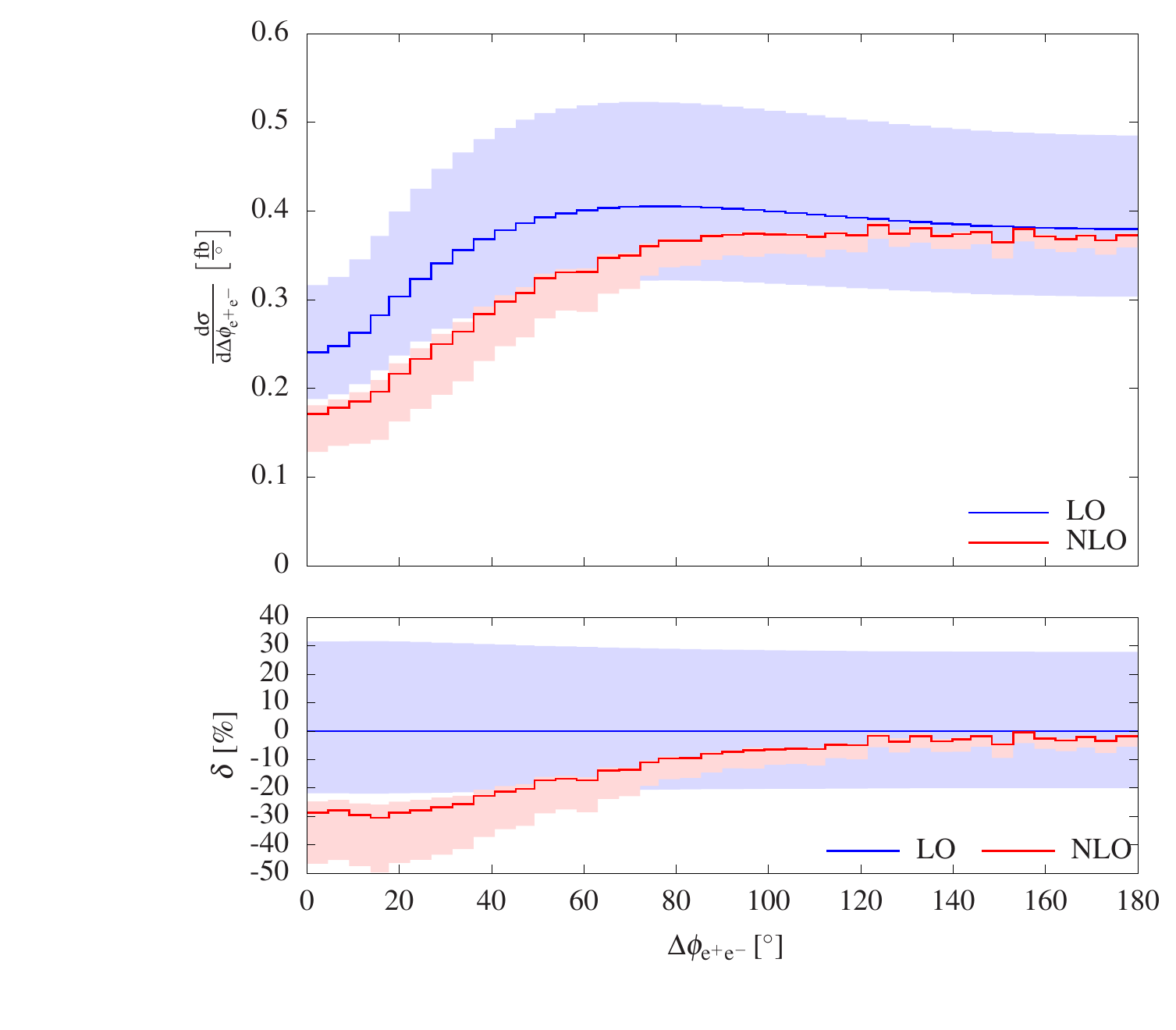}
\label{fig:aziees}
\end{subfigure}%
\par\bigskip
\begin{subfigure}{0.49\textwidth}
\centering
\subcaption{}
\includegraphics[width=1.\linewidth]{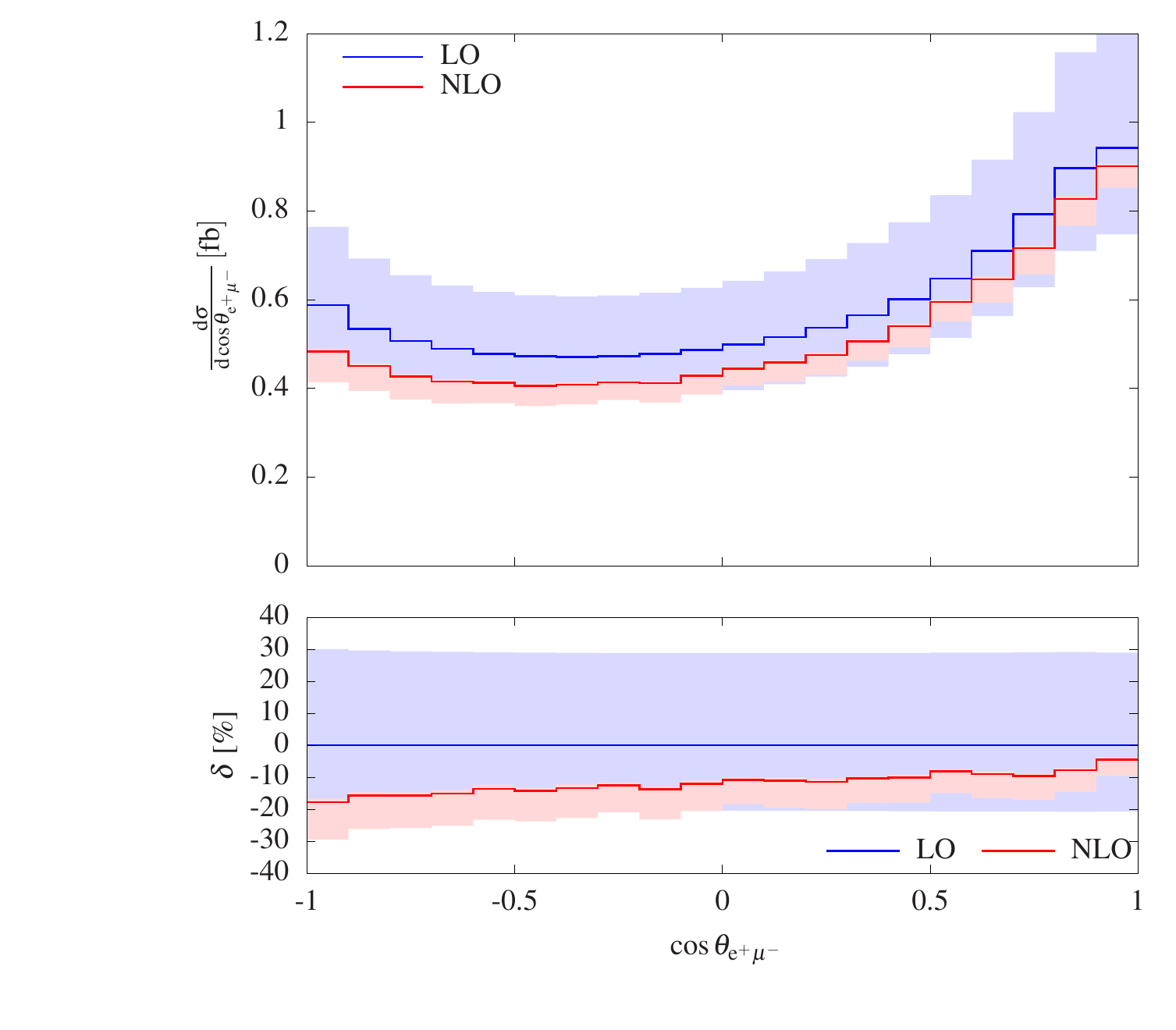}
\label{fig:cosems} 
\end{subfigure}
\begin{subfigure}{0.49\textwidth}
\centering
\subcaption{}
\includegraphics[width=1.\linewidth]{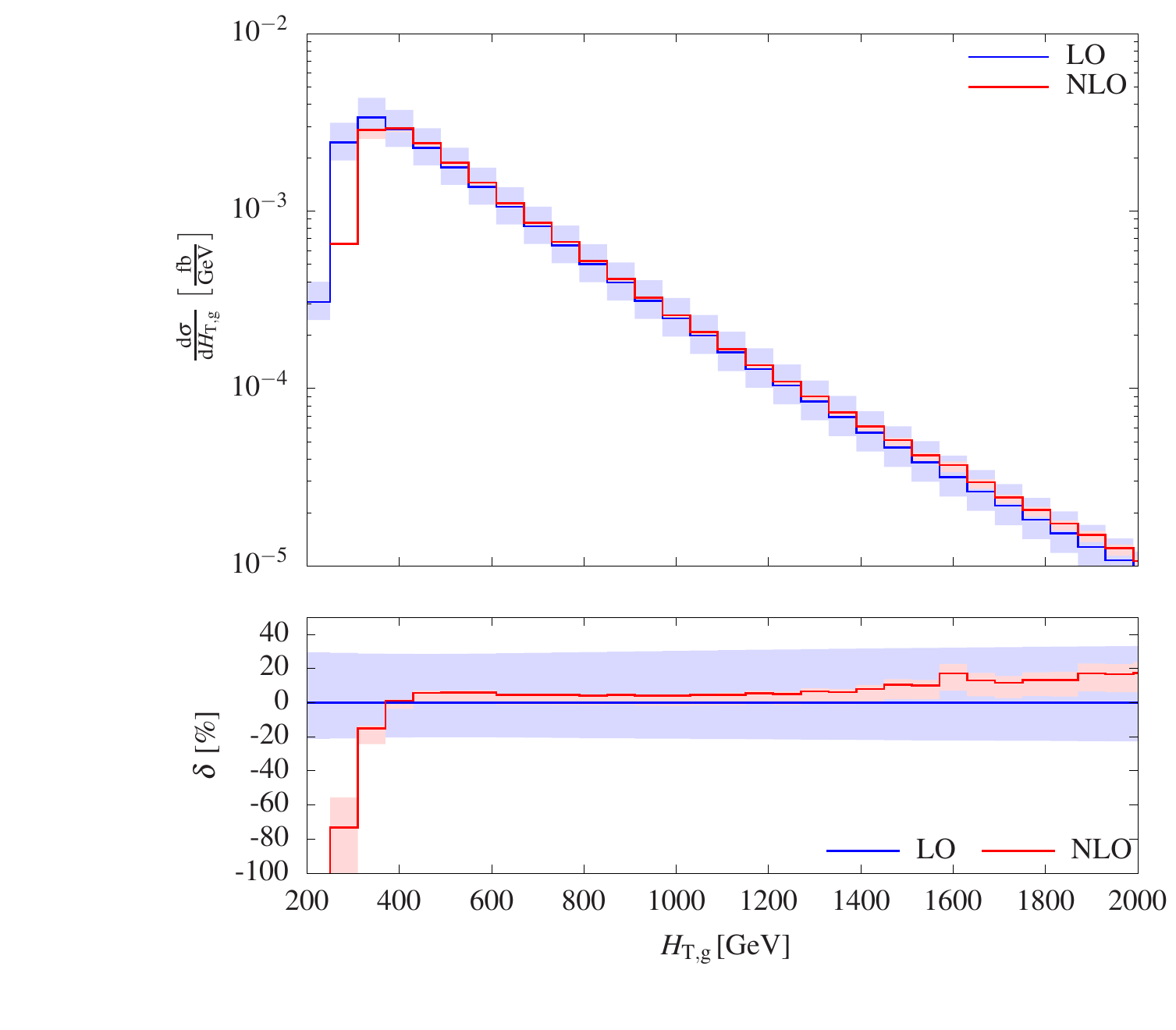}
\label{fig:Hts}
\end{subfigure}%
\vspace*{-3ex}
\caption{Same as for \reffi{fig:NLO_scale} but for the observables:
transverse momentum of the electron--positron system (top left),
azimuthal angle between the positron and the muon (top right),
cosine of the angle between the positron and the muon (bottom left), and
total transverse energy $H_{\rm T}$ (bottom right).}
\label{fig:NLO_scale2}
\end{figure}

We now turn to the discussion of scale uncertainties for
distributions.  In the upper part of \reffis{fig:NLO_scale} and \ref{fig:NLO_scale2}, absolute
LO and NLO predictions are presented including 7-point
variations of the QCD renormalisation and factorisation scales. 
The
lower parts show NLO corrections including scale uncertainties relative
to the LO predictions at the central scale  \refeq{eq:defscale} together with the relative
LO scale uncertainty.  While the full LO predictions include the
orders $\mathcal{O}(\alpha^6)$, $\mathcal{O}(\alphas\alpha^5)$, and
$\mathcal{O}(\alphas^2\alpha^4)$, the NLO ones comprise
$\mathcal{O}(\alpha^7)$, $\mathcal{O}(\alphas\alpha^6)$,
$\mathcal{O}(\alphas^2\alpha^5)$, and $\mathcal{O}(\alphas^3\alpha^4)$
contributions. In particular, these predictions do not include loop-induced
contributions of order $\order{\alphas^4 \alpha^4}$ which have already 
been shown in \citere{Denner:2020zit}.

The total corrections to the distribution in the invariant mass of the
two tagging jets (\reffi{fig:mjjs}) vary between $-10\%$ and $-30\%$
and are smallest at about $800\GeV$.  These
corrections are not dominated by one particular contribution but
result from the interplay of the four NLO contributions (see
\reffi{fig:mjj}).  The scale uncertainty is roughly of the same size
as for the fiducial cross section. On the other hand, the corrections
to the distribution in the rapidity difference of the two tagging jets
(\reffi{fig:dyjjs}) are mostly determined by the corrections of order
$\mathcal{O}(\alphas^3\alpha^4)$ and to some extent of order
$\mathcal{O}(\alphas^2\alpha^5)$.  Accidentally, the other NLO
corrections cancel each other quite well. The NLO scale uncertainty
strongly increases for extreme rapidity difference.  The NLO
corrections to the distribution in the transverse momentum of the
hardest jet (\reffi{fig:ptj1s}) range between $0\%$ and $-25\%$, where
the overall behaviour is directed by the one of the
$\mathcal{O}(\alphas^3\alpha^4)$ corrections. It is worth noting
that the NLO prediction is beyond the edge of the LO scale-uncertainty
band for $p_{\rT,\Pj_1}>200\GeV$ and that the scale uncertainty for
this distribution is rather large.  As for all other distributions,
the central value is close to the upper edge of the scale
envelope.  The behaviour of the NLO corrections to the distribution in
the transverse momentum of
the second hardest jet (\reffi{fig:ptj2s}) is quite different from
the one of the hardest jet.  The NLO corrections are at the level of
$-20\%$ at $30\GeV$ and reach a minimum at $100\GeV$ close to $0\%$ to
steadily decrease to $-20\%$ at $800\GeV$.  The NLO scale uncertainty
stays small over the whole considered kinematic range.

The NLO corrections to the distribution in the transverse momentum of
the electron--positron pair, \ie of one of the \PZ~bosons, are shown
in \reffi{fig:pTees}.  As expected from \reffi{fig:pTee}, the full NLO
corrections become negatively large towards high energy to reach
$-60\%$ at $700\GeV$.  The NLO corrections leave the LO scale
uncertainty band at about $200\GeV$, and the NLO scale uncertainty
increases significantly for high transverse momenta. This behaviour
results from our scale choice \refeq{eq:defscale}, which is tailored to
the $\order{\alpha^6}$ contributions but not to the dominating
$\order{\alphas^2\alpha^4}$ contributions at high leptonic energies,
where the scales are underestimated.
The azimuthal angle between electron and positron is correlated to the
transverse momentum of the electron--positron pair. Accordingly, the
corrections are small for $\Delta\phi_{\Pe^+\Pe^-}\approx180^\circ$
and increase smoothly to $-30\%$ for small $\Delta\phi_{\Pe^+\Pe^-}$,
where the corresponding rate is minimal and the scale uncertainty is
about $20\%$.  Since the correlation of the angle between the
positron and the muon to the transverse momentum of the
electron--positron pair is smaller, the corrections vary only from
$-5\%$ to $-20\%$ with increasing angle.  Finally, we show the
distribution in the total transverse energy, defined as the sum of
transverse energies of the tagged final-state objects,
\begin{equation}
H_\rT = E_{\rT,\Pe^+} +  E_{\rT,\Pe^-} + E_{\rT,\mu^+} +  E_{\rT,\mu^-} 
+  E_{\rT,\Pj_1} +  E_{\rT,\Pj_2},
\end{equation}
which are defined from the transverse momenta $p_{\rT,i}$ as
\begin{equation}
E_{\rT,i}=\sqrt{p_{\rT,i}^2+M_i^2},
\end{equation}
where the invariant mass $M_i$ is nonzero after recombination.
The large negative corrections below $400\GeV$ result exclusively from the
order $\order{\alphas^3\alpha^4}$ corrections, more precisely from a
suppression of the corresponding real contributions resulting from the
restriction of the related phase space for small  $H_\rT$.
The smooth increase of
the corrections to $20\%$ at $2\TeV$ is basically driven by the 
positive $\order{\alphas^3\alpha^4}$ corrections combined with
negative  $\order{\alphas^2\alpha^5}$ corrections.
The scale uncertainty remains small apart from the region of very
small $H_\rT$.

\section{Conclusion}
\label{sec:conclusion}

Given the current and upcoming expected precision of the LHC experiments, VBS processes offer great opportunities to further probe the EW sector of the Standard Model.
In this context, precision does not only refer to the EW signal that contains VBS contributions but also to the irreducible background that can be overwhelming.
In addition of being sometimes large, the background can not be trivially separated from the signal making it a crucial component of any VBS studies.
This calls for a full NLO description of VBS signatures including all possible contributions.

In this article we have followed this avenue by presenting full NLO predictions for $\Pp\Pp \to \Pe^+\Pe^-\mu^+\mu^-\Pj\Pj+X$.
While some of the NLO contributions were already known in the literature, they are shown together in a unique setup for the first time here.
This allows to single out salient features of VBS into $\PZ\PZ$ pairs.
It is worth noting that the hierarchy of NLO corrections is rather
different from the one in the ss-WW case (which was the only channel
known at full NLO accuracy up to now), in particular, owing to the
appearance of partonic channels with gluons.
Comparing two different setups, with different invariant-mass cuts on the two tagging jets, we have demonstrated that the LO hierarchy strongly influences the size of the NLO corrections.
The newly computed corrections of order $\order{\alphas^2\alpha^5}$ turn out to be much more sizeable here than for the ss-WW signature.
While for the ss-WW cross section, they were found to be below a per cent, they are between $-6\%$ and $-8\%$ (depending on the setup) for the ZZ case discussed here.
We have traced back these corrections to typical EW Sudakov logarithms
which appear for both signatures. While these corrections contribute
for VBS into $\PZ\PZ$ with their natural order of magnitude, they were 
 accidentally compensated for ss-WW by QCD corrections to the LO
 interference appearing at the same order.
This cancellation does not happen here due to the larger LO QCD contribution upon which these EW corrections act.
This has an important implication: in arbitrary experimental setups,
none of the four NLO corrections can safely be neglected at the $10\%$ level.
Indeed, the impact of the various NLO corrections strongly depends on the hierarchy between the LO contributions which eventually originates from the experimental event selections.

In high-energy tails of  distributions, the
$\order{\alphas^2\alpha^5}$ corrections reach $30\%$, while the
$\order{\alphas^3\alpha^4}$ corrections can amount to $50\%$. Angular
distributions are distorted by up to $10\%$ and $25\%$ by the
corrections of orders $\order{\alphas^2\alpha^5}$ and
$\order{\alphas^3\alpha^4}$, respectively. Owing to the enhancement
from Sudakov logarithms and the scale choice adapted to the VBS
contributions, the total NLO corrections exceed the LO scale
uncertainty band in various regions of phase space.

The results presented here should prove particularly useful for
current and upcoming analyses for $\PZ\PZ$ production in association
with two jets at the LHC.  We hope that experimental collaborations will
take into account such NLO corrections, paving the way to precise
comparisons between experimental measurements and theoretical
predictions.
We would like to emphasise that tailored corrections can be provided upon request.

\section*{Acknowledgements}

AD, RF, and TS acknowledge financial
support by the German Federal Ministry for Education and Research
(BMBF) under contract no.~05H18WWCA1 and the German Research
Foundation (DFG) under reference numbers DE 623/6-1 and DE 623/6-2.
MP acknowledges support from the German Research Foundation (DFG) through the Research Training Group RTG2044 and the state of Baden-W\"urttemberg through bwHPC.
This work received support from STSM Grants of the COST Action CA16108.

\bibliographystyle{JHEPmod}
\bibliography{vbs_zz}

\end{document}